\documentclass[aps, prresearch, reprint, longbibliography, superscriptaddress]{revtex4-2}
\usepackage[dvipdfmx]{graphicx}
\usepackage{amsmath}
\usepackage{amssymb}

\usepackage{dcolumn}
\usepackage{bm}

\usepackage[normalem]{ulem}
\usepackage{color}

\newcommand{\mb}[1]{\mathbb{#1}}

\usepackage{hyperref}

\begin{document}

\preprint{APS/123-QED}

\title{Theoretical Analysis of Resource-Induced Phase Transitions in Estimation Strategies}

\author{Takehiro Tottori}
\affiliation{Laboratory for Neural Computation and Adaptation, RIKEN Center for Brain Science, 2-1 Hirosawa, Wako, Saitama 351-0198, Japan}
\affiliation{Institute of Industrial Science, The University of Tokyo, 4-6-1 Komaba, Meguro, Tokyo 153-8505, Japan}
\author{Tetsuya J. Kobayashi}
\affiliation{Institute of Industrial Science, The University of Tokyo, 4-6-1 Komaba, Meguro, Tokyo 153-8505, Japan}

\date{\today}

\begin{abstract}
Organisms adapt to volatile environments by integrating sensory information with internal memory, yet their information processing is constrained by resource limitations. 
Such limitations can fundamentally alter optimal estimation strategies in biological systems. 
For example, recent experiments suggest that organisms exhibit nonmonotonic phase transitions between memoryless and memory-based estimation strategies depending on sensory reliability. 
However, an analytical understanding of these resource-induced phase transitions is still missing. 
This Letter presents an analytical characterization of resource-induced phase transitions in optimal estimation strategies. 
Our result identifies the conditions under which resource limitations alter estimation strategies and analytically reveals the mechanism underlying the emergence of discontinuous, nonmonotonic, and scaling behaviors. 
These results provide a theoretical foundation for understanding how limited resources shape information processing in biological systems.
\end{abstract}

\maketitle

\begin{figure*}
	\includegraphics[width=170mm]{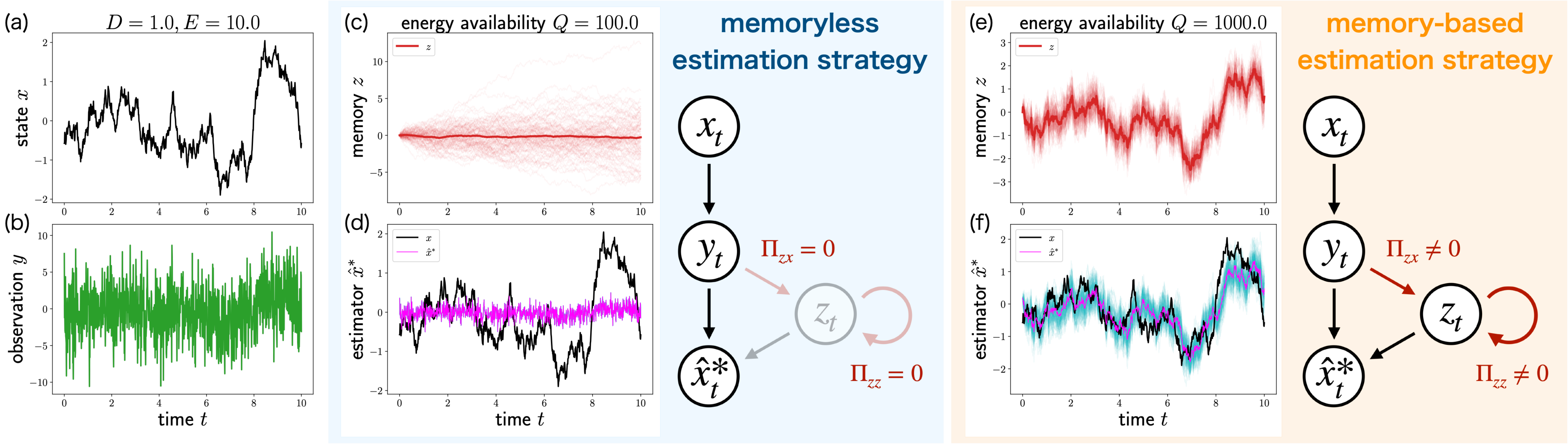}
	\caption{\label{fig-1}
	(a) Environmental state $x_{t}$. 
	(b) Noisy observation $y_{t}$. 
	(c,e) Internal memory $z_{t}$. 
	(d,f) Optimal estimator $\hat{x}_{t}^{*}$. 
	Thin red and cyan curves are 100 sample trajectories, whereas thick red and magenta curves are their means. 
	While memoryless estimation strategy ($\Pi_{zx},\Pi_{zz}=0$) is optimal at $Q=100$ (c,d), memory-based one ($\Pi_{zx},\Pi_{zz}\neq0$) is optimal at $Q=1000$ (e,f). 
	The rest of the parameters are set to $F=1$ and $M=1$.
	}
\end{figure*}
\begin{figure*}
	\includegraphics[width=170mm]{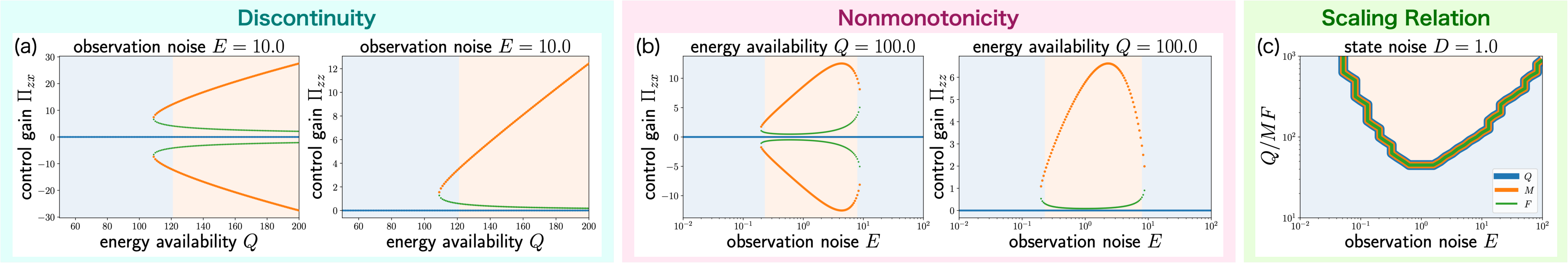}
	\caption{\label{fig-2}
	(a,b) $\Pi_{zx}$ and $\Pi_{zz}$ as functions of $Q$ and $E$. 
	Blue, green, and orange dots are the numerical solutions of the observation-based Riccati equation, and correspond to zero, intermediate, and high memory control gains, respectively. 
	The blue and orange dots are optimal in the blue and orange regions, respectively, whereas the green dots never become optimal. 
	(c) Phase boundaries with respect to $E$ and $Q/MF$. 
	Blue, green, and orange curves vary $Q$, $M$, and $F$, respectively. 
	The rest of the parameters are set to $1$. 
	}
\end{figure*}

{\it Introduction.--}
Organisms adapt to changing environments by integrating sensory information with internal memory \cite{kobayashi_implementation_2010,hinczewski_cellular_2014,becker_optimal_2015,mora_physical_2019,nakamura_connection_2021,heinonen_optimal_2023,rode_information_2024,krishnamurthy_arousal-related_2017,piet_rats_2018,kutschireiter_bayesian_2023}. 
However, the resources they can devote to information processing are intrinsically limited \cite{lan_energyspeedaccuracy_2012,govern_energy_2014,govern_optimal_2014,lang_thermodynamics_2014,ito_maxwells_2015,tjalma_trade-offs_2023,bryant_physical_2023,nicoletti_tuning_2024,laughlin_metabolic_1998,lennie_cost_2003,burns_costs_2010,lieder_resource-rational_2020,polania_rationality_2024,tavoni_what_2019,bhui_resource-rational_2021}, which in turn shapes their optimal estimation strategies. 
For example, when energy resources are sufficient, organisms can engage in deliberative decision-making, whereas under severe resource limitations their behavior tends to become more reactive \cite{kahneman_thinking_2011,gailliot_self-control_2007,masicampo_toward_2008}. 
Moreover, recent experiments suggest that although organisms exploit their internal memory when sensory information is moderately uncertain, they abandon it when sensory uncertainty is either too low or too high \cite{tavoni_human_2022}. 
Despite these observations, a theoretical framework that explains when and how resource limitations induce phase transitions in optimal estimation strategies remains missing. 

By employing optimal control theory, we previously developed a general framework for identifying optimal estimation strategies under resource limitations \cite{tottori_memory-limited_2022,tottori_forward-backward_2023,tottori_decentralized_2023,tottori_resource_2025,tottori_theory_2025}. 
By applying this framework to a minimal model of biological information processing, we found that resource limitations induce a phase transition between memoryless and memory-based estimation strategies [Fig. \ref{fig-1}] \cite{tottori_resource_2025,tottori_theory_2025}. 
This transition also exhibits several nontrivial features, including discontinuity, nonmonotonicity, and a scaling relation, even in a linear-quadratic-Gaussian (LQG) setting [Fig. \ref{fig-2}] \cite{tottori_resource_2025,tottori_theory_2025}. 
The observed nonmonotonicity qualitatively agrees with recent experimental results \cite{tavoni_human_2022}, such that memory-based estimation is optimal only when sensory uncertainty is moderate. 
However, these findings were obtained from numerical simulations, and their mechanism is yet to be clarified. 

This letter elucidates the mechanism underlying the resource-induced phase transition. 
Our analysis clarifies when and how resource limitations induce phase transitions in optimal estimation strategies, and reveals their discontinuous, nonmonotonic, and scaling behaviors analytically. 
These results provide a theoretical foundation for understanding resource-constrained information processing in biological systems. 

{\it Model.--}
We formulate a minimal model of biological information processing, in which an organism estimates an environmental state $x_{t}\in\mathbb{R}$ by integrating noisy observation $y_{t}\in\mathbb{R}$ and internal memory $z_{t}\in\mathbb{R}$ \cite{tottori_resource_2025,tottori_theory_2025}. 

The environmental state $x_{t}\in\mathbb{R}$ evolves according to an Ornstein-Uhlenbeck process [Fig. \ref{fig-1}(a)] \cite{hinczewski_cellular_2014,becker_optimal_2015,tjalma_trade-offs_2023}:
\begin{align}
	dx_{t}=-x_{t}dt+\sqrt{D}d\omega_{t},
	\label{eq: state}
\end{align}
where $\omega_{t}$ is a standard Wiener process and $D > 0$ denotes the intensity of state fluctuations. 
The organism cannot observe $x_{t}$ directly but receives a noisy measurement $y_{t} \in \mathbb{R}$ drawn from a Gaussian distribution with mean $x_{t}$ and variance $E>0$ [Fig. \ref{fig-1}(b)]:
\begin{align}
	y_{t}\sim\mathcal{N}(y_{t}|x_{t},E).
	\label{eq: observation}
\end{align}
To utilize past information, the organism maintains an internal memory $z_{t}\in\mathbb{R}$, which is updated as
\begin{align}
	dz_{t}=v_{t}dt+\sqrt{F}d\xi_{t},
	\label{eq: memory}
\end{align}
where $\xi_{t}$ is a standard Wiener process and $F>0$ denotes the intensity of intrinsic noise [Fig. \ref{fig-1}(c,e)]. 
The control function $v_{t}=v(y_{t}, z_{t})$ determines how the organism encodes past observations into memory. 
The organism estimates the environmental state $x_{t}$ by integrating the current observation $y_{t}$ with the internal memory $z_{t}$ as $\hat{x}_{t}=\hat{x}(y_{t},z_{t})$ [Fig. \ref{fig-1}(d,f)]. 

The optimal state estimator $\hat{x}^{*}$ and memory control $v^{*}$ are defined as the minimizers of the following objective: 
\begin{align}
	&\hat{x}^{*},v^{*}:=\arg\min_{\hat{x},v}J[\hat{x},v],\\
	&J[\hat{x}, v]:= \lim_{T \to \infty} \frac{1}{T} \mathbb{E} \left[ \int_0^T \left( Q(x_{t} - \hat{x}_{t})^2 + M v_{t}^{2} \right) dt \right]. 
	\label{eq: objective}
\end{align}
The first and second terms in Eq. (\ref{eq: objective}) represent the state estimation error and the memory control cost, respectively. 
The parameters $Q>0$ and $M>0$ quantify energy availability and limitation, respectively. 
Increasing $Q$ or decreasing $M$ allows the organism to invest more energy in estimation. 
This problem corresponds to the LQG problem in optimal control theory \cite{yong_stochastic_1999,nisio_stochastic_2015,bensoussan_estimation_2018}, since the dynamics  are linear and Gaussian [Eqs. (\ref{eq: state})--(\ref{eq: memory})] and the cost is quadratic [Eq. (\ref{eq: objective})]. 

By employing optimal control theory, the solution satisfying the stationary condition of $J[\hat{x},v]$ is given by
\begin{align}
	\hat{x}^{*}(y,z)&= \mathbb{E}_{p(x|y, z)}[x]=K_{xy}y+K_{xz}z,\label{eq: optimal estimator}\\
	v^{*}(y,z)&=-M^{-1}\Pi_{zx}\hat{x}^{*}(y,z)-M^{-1}\Pi_{zz}z,\label{eq: optimal control}
\end{align}
where the estimation gains $K_{xy}$ and $K_{xz}$ and control gains $\Pi_{zx}$ and $\Pi_{zz}$ are the solutions of the observation-based Riccati equation [see Sec. I in Supplemental Material] \cite{tottori_resource_2025,tottori_theory_2025}. 
However, this equation is multivariable and nonlinear, making analytical treatment intractable. 
Therefore, the previous work investigated it numerically \cite{tottori_resource_2025,tottori_theory_2025}. 

{\it Numerical Results.--}
Before presenting the analytical results, we briefly summarize the numerical results of Eqs. (\ref{eq: optimal estimator}) and (\ref{eq: optimal control}) \cite{tottori_resource_2025,tottori_theory_2025}. 
At low energy availability ($Q=100$), the memory control gains, $\Pi_{zx}$ and $\Pi_{zz}$, vanish, indicating that no observational information is encoded into the memory $z_{t}$ [Fig. \ref{fig-1}(c,d)]. 
As a result, the organism estimates the environmental state $x_{t}$ solely based on the current observation $y_{t}$, corresponding to memoryless estimation. 
In contrast, when more energy is available ($Q=1000$), $\Pi_{zx}$ and $\Pi_{zz}$ take nonzero values, and memory-based estimation becomes optimal [Fig. \ref{fig-1}(e,f)]. 
These results demonstrate a qualitative change in estimation strategies depending on energy availability. 

This phase transition is discontinuous even though this model is linear and Gaussian [Fig. \ref{fig-2}(a)]. 
At low $Q$, $\Pi_{zx}=\Pi_{zz}=0$ is the only solution and clearly optimal. 
As $Q$ increases, $\Pi_{zx},\Pi_{zz}\neq0$ emerge discontinuously, while $\Pi_{zx}=\Pi_{zz}=0$ remains optimal. 
After further increase of $Q$, the optimal solution discontinuously switches from $\Pi_{zx}=\Pi_{zz}=0$ to $\Pi_{zx},\Pi_{zz}\neq0$. 
This indicates that estimation strategies change qualitatively with energy availability. 

While increasing $Q$ induces a monotonic transition from memoryless to memory-based estimation [Fig. \ref{fig-2}(a)], increasing the observation noise $E$ leads to a nonmonotonic transition [Fig. \ref{fig-2}(b)]. 
At low $E$, memoryless estimation is optimal since the current observation contains sufficient information. 
As $E$ increases, the current observation becomes less reliable, and memory-based estimation becomes favorable. 
However, when $E$ becomes too large, memory control becomes ineffective, and memoryless estimation again becomes optimal. 
This nonmonotonic phase transition is also observed in recent human experiments \cite{tavoni_human_2022}. 
Furthermore, the state noise $D$ also induces a similar nonmonotonic phase transition \cite{tottori_resource_2025,tottori_theory_2025}. 

Similar to the transition with respect to $Q$, those with respect to $M$ and $F$ also occur monotonically but in the opposite direction: 
increasing $M$ or $F$ induces a transitions from memory-based to memoryless estimation \cite{tottori_resource_2025,tottori_theory_2025}. 
Moreover, the phase boundaries with respect to $Q$, $M$, and $F$ collapse onto a single curve when plotted against $Q/MF$ [Fig. \ref{fig-2}(c)], indicating a scaling relation among these parameters. 
While the scaling relation $Q/M$ is trivial, the scaling relation $Q/MF$ is not because $Q$ and $M$ represent energy availability, whereas $F$ characterizes intrinsic noise. 

{\it Theoretical Analysis.--}
In this Letter, we analytically clarify the mechanism of these numerically observed phenomena in the resource-induced phase transitions. 
The main obstacle hindering analytical treatment is the complexity of the observation-based Riccati equation, which must be solved to determine the estimation gains $K_{xy}$ and $K_{xz}$ and the control gains $\Pi_{zx}$ and $\Pi_{zz}$ [Eqs. (\ref{eq: optimal estimator}) and (\ref{eq: optimal control})]. 
To overcome this difficulty, we reformulate the memory control function as a linear function of the observation $y$ and memory $z$: 
\begin{align}
	v_{\Phi}(y,z):=-\Phi_{zy}y-\Phi_{zz}z, 
	\label{eq: linear control function}
\end{align}
where $\Phi_{zy}\in\mb{R}$ and $\Phi_{zz}\geq0$. 
$\Phi_{zy}$ is the control gain for encoding observational information, whereas $\Phi_{zz}$ provides negative feedback to suppress memory fluctuations. 
This reformulation does not compromise optimality because the optimal memory control function $v^{*}(y,z)$ is also linear  [Eq. (\ref{eq: optimal control})]. 
Moreover, this reformulation enables us to derive an analytical expression of the objective function $J$ [see Sec. II in Supplemental Material]. 

\begin{figure}
	\includegraphics[width=85mm]{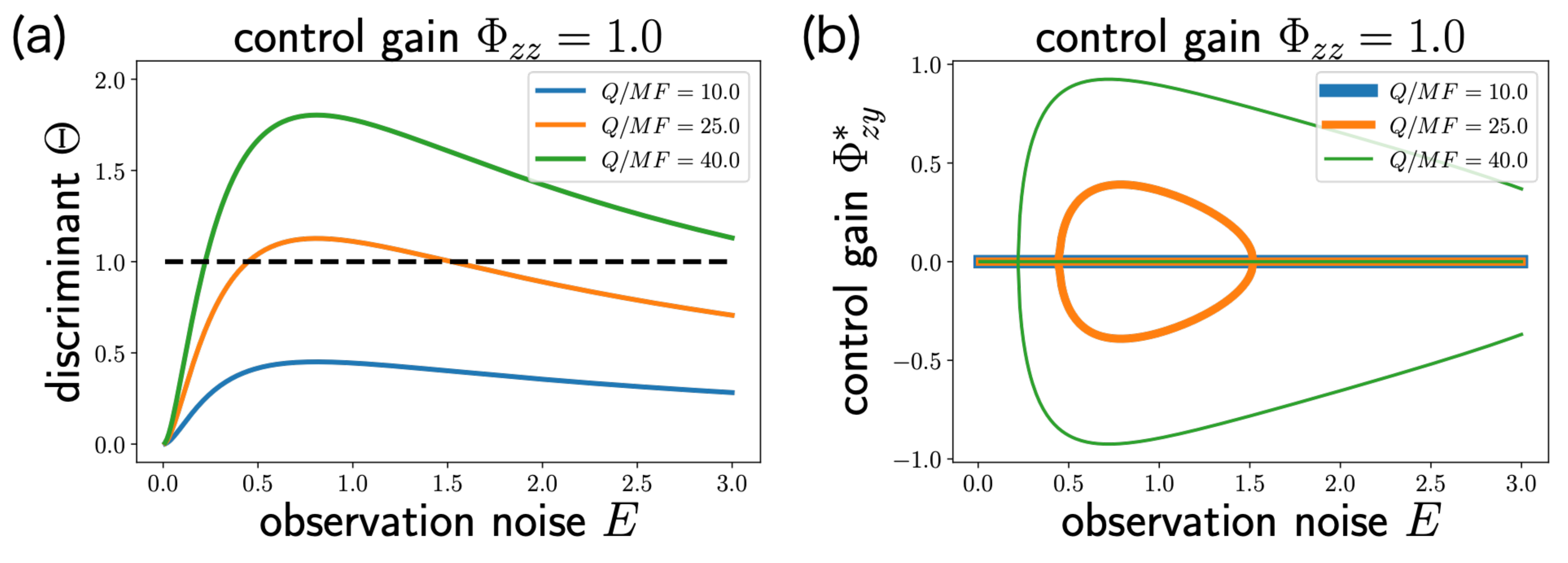}
	\caption{\label{fig-3}
	Discriminant $\Theta$ [Eq. (\ref{eq: discriminant 1})] and optimal memory control gain $\Phi_{zy}^{*}$ [Eq. (\ref{eq: optimal Pzy 1})] for $\Phi_{zz}=1$. 
	$D$ is set to 1. 
	}
\end{figure}

First, we fix $\Phi_{zz}$ and optimize only $\Phi_{zy}$. 
From the stationarity condition $\partial J/\partial \Phi_{zy}=0$, we obtain the following optimal control gain [see Sec III. in Supplemental Material]:
\begin{align}
	&\Phi_{zy}^{*}=0,\ \pm\sqrt{\frac{\left(\Phi_{zz}+1\right)^{2}\left(D+2E\right)F}{\left\{D+2\left(\Phi_{zz}+1\right)E\right\}D}\left(\sqrt{\Theta}-1\right)},
	\label{eq: optimal Pzy 1}\\
	&\Theta:=\frac{4\Phi_{zz}D^{2}E^{2}\left(Q/MF\right)}{\left(\Phi_{zz}+1\right)\left\{D+2\left(\Phi_{zz}+1\right)E\right\}\left(D+2E\right)^{2}}.
	\label{eq: discriminant 1}
\end{align}
When $\Theta\leq1$, the only real solution is $\Phi_{zy}^{*}=0$, meaning that no observational information is encoded into memory. 
Thus, the memoryless estimation strategy is optimal. 
On the other hand, when $\Theta>1$, nonzero real solutions $\Phi_{zy}^{*}\neq0$ emerge, indicating that memory-based estimation can be optimal. 
Therefore, $\Theta$ serves as a discriminant identifying the phase boundary between memoryless and memory-based estimation. 

$\Theta$ captures the scaling relation $Q/MF$ because it depends on $Q$, $M$, and $F$ only through this form [Eq. (\ref{eq: discriminant 1})]. 
$\Theta$ increases monotonically with increasing $Q/MF$, which accounts for the monotonic phase transitions with respect to $Q$, $M$, and $F$. 

In contrast, the dependence of $\Theta$ on $E$ and  $D$ is more intricate [Fig. \ref{fig-3}(a)]. 
When $E\ll D$, $\Theta\approx4\left(Q/MF\right)\left(\Phi_{zz}+1\right)^{-1}\Phi_{zz}D^{-1}E^{2}$, which increases with increasing $E$ or decreasing $D$. 
Conversely, when $E\gg D$, $\Theta\approx\left(Q/MF\right)\left(\Phi_{zz}+1\right)^{-2}\Phi_{zz}D^{2}E^{-1}$, which decreases with increasing $E$ or decreasing $D$. 
Therefore, $\Theta$ also accounts for the nonmonotonic phase transitions with respect to $E$ and $D$. 

\begin{figure}
	\includegraphics[width=85mm]{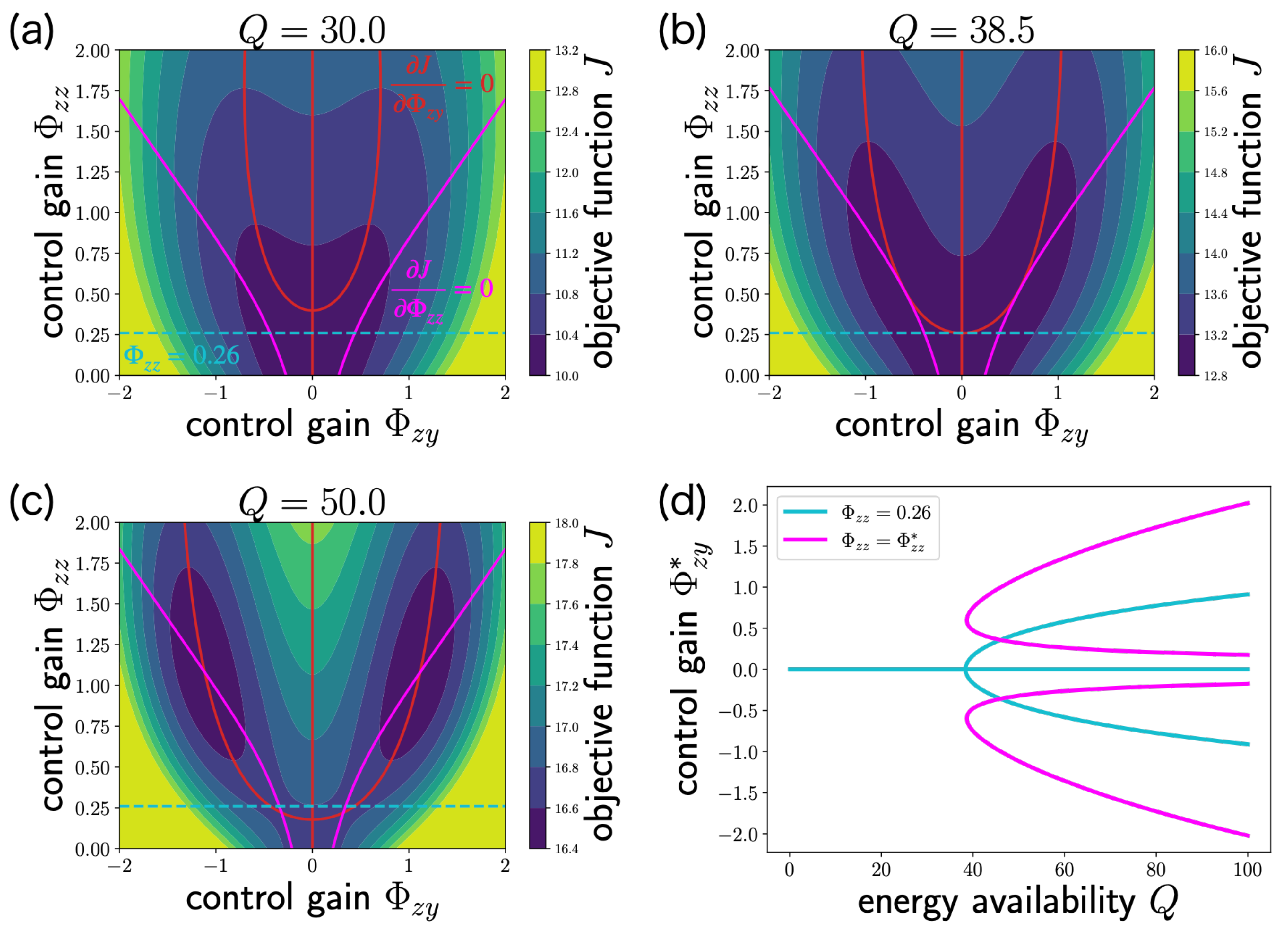}
	\caption{\label{fig-4}
	(a,b,c) $J$ as functions of $\Phi_{zy}$ and $\Phi_{zz}$. 
	Red and magenta curves represent $\partial J/\partial \Phi_{zy}=0$ and $\partial J/\partial \Phi_{zz}=0$, respectively, while cyan lines indicate $\Phi_{zz}=0.26$. 
	(d) $\Phi_{zy}^{*}$ as a function of $Q$. 
	Cyan curves are the intersections between $\partial J/\partial \Phi_{zy}=0$ and $\Phi_{zz}=0.26$, whereas magenta curves correspond to the intersections between $\partial J/\partial \Phi_{zy}=0$ and $\partial J/\partial \Phi_{zz}=0$. 
	The rest of the parameters are set to 1. 
	}
\end{figure}

However, Eqs. (\ref{eq: optimal Pzy 1}) and (\ref{eq: discriminant 1}) do not capture the discontinuity of the phase transition, since $\Phi_{zy}^{*}$ varies continuously from zero to nonzero values [Fig. \ref{fig-3}(b)]. 
In fact, the intersections between $\partial J/\partial \Phi_{zy}=0$ [Fig. \ref{fig-4}(a-c), red] and $\Phi_{zz}={\rm constant}$ [Fig. \ref{fig-4}(a-c), cyan] change continuously from zero to nonzero values as $Q$ increases [Fig. \ref{fig-4}(d), cyan]. 
This result suggests that the simultaneous optimization of both $\Phi_{zy}$ and $\Phi_{zz}$ is essential to produce the discontinuous phase transition. 
Indeed, the intersections between $\partial J/\partial \Phi_{zy}=0$ [Fig. \ref{fig-4}(a-c), red] and $\partial J/\partial \Phi_{zz}=0$ [Fig. \ref{fig-4}(a-c), magenta] emerge discontinuously at nonzero values as $Q$ increases [Fig. \ref{fig-4}(d), magenta]. 
These results indicate that the discontinuity arises from the simultaneous optimization of two distinct memory control gains: $\Phi_{zy}$, which encodes the observational information, and $\Phi_{zz}$, which stabilizes the memory state. 

\begin{figure}
	\includegraphics[width=85mm]{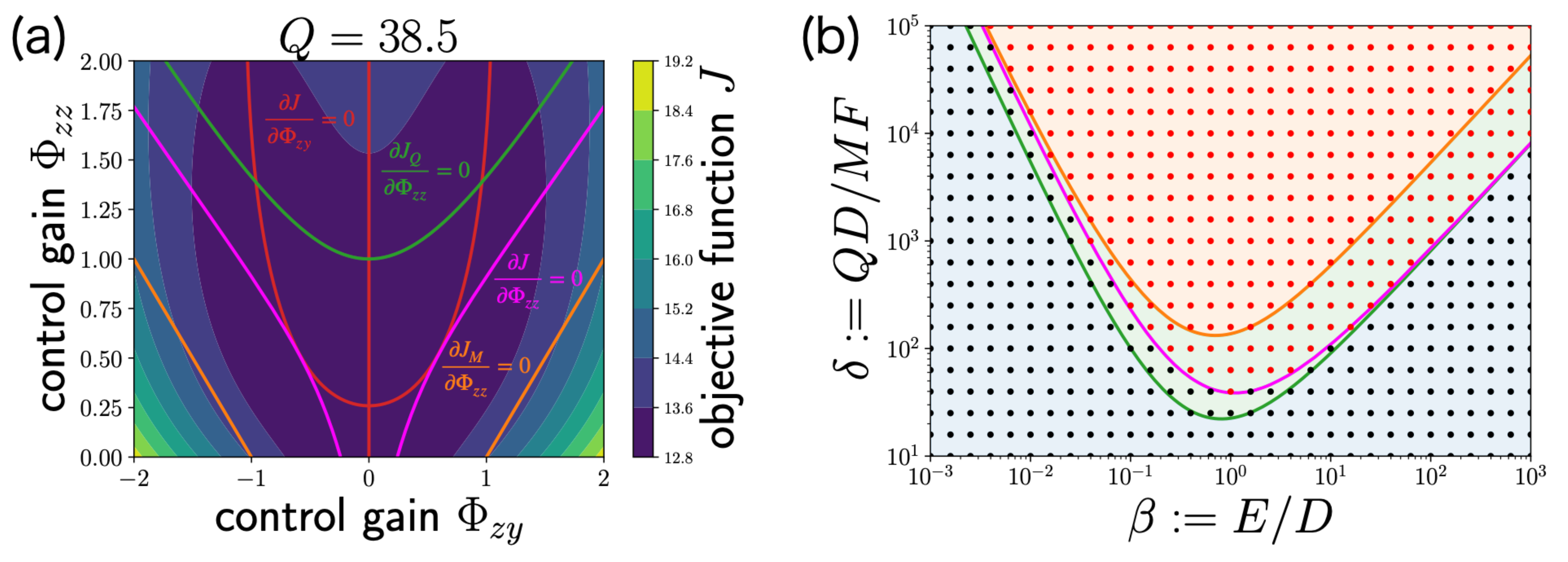}
	\caption{\label{fig-5}
	(a) Red, magenta, green, and orange curves indicate $\partial J/\partial \Phi_{zy}=0$, $\partial J/\partial \Phi_{zz}=0$, $\partial J_{Q}/\partial \Phi_{zz}=0$, and $\partial J_{M}/\partial \Phi_{zz}=0$, respectively. 
	The rest of the parameters are set to 1. 
	(b) Magenta, green, and orange curves are the parameter values at which $\partial J/\partial \Phi_{zz}=0$, $\partial J_{Q}/\partial \Phi_{zz}=0$, and $\partial J_{M}/\partial \Phi_{zz}=0$ intersect tangentially with $\partial J/\partial \Phi_{zy}=0$, respectively. 
	The magenta curve ($\Theta_{T}=1$) is obtained numerically, whereas the green ($\Theta_{Q}=1$) and orange ($\Theta_{M}=1$) curves are obtained analytically from Eqs. (\ref{eq: main-theta-q}) and (\ref{eq: main-theta-m}), respectively. 
	Blue, green, and orange regions correspond to $\Theta_{Q}<1$, $1\leq\Theta_{Q}$ and $\Theta_{M}<1$, and $1\leq\Theta_{M}$, respectively. 
	The control gains $\Pi_{zx}$ and $\Pi_{zz}$ obtained numerically from the observation-based Riccati equation are zero at black dots and take nonzero values at red dots. 
	}
\end{figure}

We next investigate the simultaneous optimization of $\Phi_{zy}$ and $\Phi_{zz}$ more analytically. 
While $\partial J/\partial \Phi_{zy}=0$ can be solved analytically, $\partial J/\partial \Phi_{zz}=0$ is intractable analytically. 
To address this issue, we decompose the objective function $J$ into the state estimation error $J_{Q}$ and the memory control cost $J_{M}$ as follows: 
\begin{align}
	&J=J_{Q}+J_{M},\\
	&J_{Q}:= \lim_{T \to \infty} \frac{1}{T} \mathbb{E} \left[ \int_0^T Q(x_{t} - \hat{x}_{t})^2 dt \right],\\
	&J_{M}:= \lim_{T \to \infty} \frac{1}{T} \mathbb{E} \left[ \int_0^T M v_{t}^{2} dt \right].
\end{align}
Although $\partial J/\partial \Phi_{zz}=0$ is intractable, its components $\partial J_{Q}/\partial \Phi_{zz}=0$ and $\partial J_{M}/\partial \Phi_{zz}=0$ are more tractable. 
Moreover, the values of $\Phi_{zz}$ that satisfy $\partial J_{Q}/\partial \Phi_{zz}=0$ [Fig. \ref{fig-5}(a), green] and $\partial J_{M}/\partial \Phi_{zz}=0$ [Fig. \ref{fig-5}(a), orange] serve as the upper and lower bounds, respectively, for the value of $\Phi_{zz}$ that satisfies $\partial J/\partial \Phi_{zz}=0$ [Fig. \ref{fig-5}(a), magenta] [see Sec. IV B in Supplemental Material]. 
Thus, analyzing $\partial J_{Q}/\partial \Phi_{zz}=0$ and $\partial J_{M}/\partial \Phi_{zz}=0$ enables us to identify the necessary and sufficient conditions for the intersection between $\partial J/\partial \Phi_{zy}=0$ and $\partial J/\partial \Phi_{zz}=0$. 

The conditions under which $\partial J_{Q}/\partial \Phi_{zz}=0$ and $\partial J_{M}/\partial \Phi_{zz}=0$ intersect with $\partial J/\partial \Phi_{zy}=0$ are given by $\Theta_{Q}\geq1$ and $\Theta_{M}\geq1$, respectively, where 
\begin{align}
	&\Theta_{Q}=\frac{2\beta^{2}\delta}{\left(1+2\beta\right)^{2}\left(1+4\beta\right)},\label{eq: main-theta-q}\\
	&\Theta_{M}\approx\frac{\beta^{2}\delta}{\left(1+2\beta\right)^{2}\left\{\left(1+6\beta\right)+\sqrt{4\beta\left(4+13\beta\right)}\right\}},\label{eq: main-theta-m}
\end{align}
with $\beta:=E/D$ and $\delta:=QD/MF$ [see Sec. IV C and D in Supplemental Material]. 
These expressions suggest that the key parameters governing the resource-induced phase transitions are only $\beta$ and $\delta$: $\beta$ represents sensory uncertainty, and $\delta$ represents energy availability or memory accuracy. 
This observation can be analytically demonstrated without approximation [see Sec. IV A in Supplemental Material].  
Since $\delta$ includes $Q/MF$, it captures the scaling relation. 
In addition, while $\Theta_{Q}$ and $\Theta_{M}$ vary monotonically with $\delta$, they exhibit nonmonotonic dependence on $\beta$ [Fig. \ref{fig-5}(b), green and orange curves], thereby reproducing the nonmonotonic behavior. 

We then define the condition under which $\partial J/\partial \Phi_{zz}=0$ intersects with $\partial J/\partial \Phi_{zy}=0$ as $\Theta_{T}\geq1$. 
Although $\Theta_{T}$ is analytically intractable, $\Theta_{Q}\geq\Theta_{T}\geq\Theta_{M}$ always holds [Fig. \ref{fig-5}(b), green, magenta, and orange curves], because the values of $\Phi_{zz}$ that satisfy $\partial J_{Q}/\partial \Phi_{zz}=0$ [Fig. \ref{fig-5}(a), green] and $\partial J_{M}/\partial \Phi_{zz}=0$ [Fig. \ref{fig-5}(a), orange] serve as the upper and lower bounds, respectively, for the value of $\Phi_{zz}$ that satisfies $\partial J/\partial \Phi_{zz}=0$ [Fig. \ref{fig-5}(a), magenta] [see Sec. IV B in Supplemental Material]. 
When $\Theta_{Q}<1$ [Fig. \ref{fig-5}(b), blue region], $\Theta_{T}<1$ always holds, and $\partial J/\partial \Phi_{zy}=0$ and $\partial J/\partial \Phi_{zz}=0$ do not intersect. 
Thus, nonzero memory control gains do not emerge [Fig. \ref{fig-5}(b), black dots]. 
In contrast, when $\Theta_{M}\geq1$ [Fig. \ref{fig-5}(b), orange region], $\Theta_{T}\geq1$ always holds, and $\partial J/\partial \Phi_{zy}=0$ and $\partial J/\partial \Phi_{zz}=0$ intersect. 
Consequently, nonzero memory control gains emerge [Fig. \ref{fig-5}(b), red dots]. 
Finally, when $\Theta_{Q}\geq1$ and $\Theta_{M}<1$ [Fig. \ref{fig-5}(b), green region], it remains ambiguous whether nonzero memory control gains emerge or not [Fig. \ref{fig-5}(b), black and red dots]. 
Since this analytical classification [Fig. \ref{fig-5}(b), blue, green, and orange regions] is consistent with the numerical result [Fig. \ref{fig-5}(b), black and red dots],  we thereby establish an analytical foundation for the resource-induced phase transitions. 

{\it Discussion.--}
In this Letter, we analytically clarified the mechanism for the discontinuous, nonmonotonic, and scaling behaviors in the resource-induced phase transitions. 
Discontinuous phase transitions in information-processing strategies have also been reported within the information bottleneck framework \cite{tishby_information_2000,chechik_information_2005,creutzig_past-future_2009,sachdeva_optimal_2021,galstyan_intuitive_2025}. 
While the information bottleneck formulates resource limitations in terms of mutual information, our optimal control framework expresses them through control costs \cite{tottori_memory-limited_2022,tottori_forward-backward_2023,tottori_decentralized_2023,tottori_resource_2025,tottori_theory_2025}. 
Despite these modeling differences, both frameworks exhibit discontinuous phase transitions even in linear-Gaussian settings, suggesting that such discontinuities could be characteristic of resource-limited information processing. 

In addition, our work analytically revealed nonmonotonic phase transitions with respect to sensory uncertainty. 
Similar nonmonotonic transitions have been reported by recent human experiments \cite{tavoni_human_2022}. 
These experiments have also reported that increasing environmental volatility induces a monotonic phase transition from memory-based to memoryless estimation strategies \cite{tavoni_human_2022}. 
Our framework analytically reproduces this behavior as well [see Sec. V in Supplemental Material]. 
These results suggest that human information processing may also be characterized by resource limitations. 

Memory-based information processing [Fig. \ref{fig-1}(e,f)] could correspond to coherent feed-forward loops (cFFLs) \cite{alon_introduction_2019,alon_network_2007,mangan_coherent_2003,mangan_structure_2003,dekel_environmental_2005}.  
A cFFL consists of a fast direct pathway and a slow indirect pathway, which filters out short-lived and spurious signals by integrating these inputs. 
Such motifs are known to be highly overrepresented in transcriptional, signaling, and neural networks compared with random networks \cite{milo_network_2002,shen-orr_network_2002,milo_superfamilies_2004}.
Our framework may offer a statistical perspective on when and how cFFLs become more adaptive than simpler direct pathways. 

The first author was supported by JST ACT-X (Grant Number JPMJAX24LB) and Special Postdoctoral Researcher (SPDR) Program at RIKEN. 
This research was supported by JST CREST (Grant Number JPMJCR2011) and JSPS KAKENHI (Grant Number 25H01365).

\bibliographystyle{apsrev4-2}
\nocite{apsrev42Control}
\bibliography{251015_OMSC_ref_cleaned}

\end{document}


\preprint{APS/123-QED}

\onecolumngrid

\vspace*{2em}
\begin{center}
    \large \textbf{Supplemental Material for:\\
    Theoretical Analysis of Resource-Induced Phase Transitions in Estimation Strategies}

    \vspace{1em}
    \normalsize
    Takehiro Tottori$^{1,2}$ and Tetsuya J. Kobayashi$^2$ \\
    \vspace{0.5em}
    \textit{
    $^{1}$Laboratory for Neural Computation and Adaptation, RIKEN Center for Brain Science,\\ 2-1 Hirosawa, Wako, Saitama 351-0198, Japan\\
    $^{2}$Institute of Industrial Science, The University of Tokyo, 4-6-1 Komaba, Meguro, Tokyo 153-8505, Japan}
\end{center}

\vspace{2em}

\setcounter{tocdepth}{1}
\tableofcontents


\setcounter{equation}{0}
\renewcommand{\theequation}{S\arabic{equation}}
\setcounter{figure}{0}
\renewcommand{\thefigure}{S\arabic{figure}}

\section{Review of Previous Work}
In this section, we briefly review the optimality condition derived in the previous work \cite{tottori_resource_2025,tottori_theory_2025}. 
The previous work introduced the following extended variables: extended state $\tilde{x}_{t}$, extended observation $\tilde{y}_{t}$, extended control $\tilde{u}_{t}$, and extended standard Wiener process $\tilde{\omega}_{t}$.  
\begin{align}
	\tilde{x}_{t}:=\left(\begin{array}{c}
		x_{t}\\
		z_{t}\\
	\end{array}
	\right),\quad 
	\tilde{y}_{t}:=\left(\begin{array}{c}
		y_{t}\\
		z_{t}\\
	\end{array}
	\right),\quad 
	\tilde{u}_{t}:=\left(\begin{array}{c}
		\hat{x}_{t}\\
		v_{t}\\
	\end{array}
	\right),\quad
	\tilde{\omega}_{t}:=\left(\begin{array}{c}
		\omega_{t}\\
		\xi_{t}\\
	\end{array}
	\right).
\end{align}
By using these extended variables, the minimal model described in the main text can be transformed into the following linear-quadratic-Gaussian (LQG) problem in optimal control theory: 
\begin{align}
	&d\tilde{x}_{t}=\left(\tilde{A}\tilde{x}_{t}+\tilde{B}\tilde{u}_{t}\right)dt+\tilde{\sigma}d\tilde{\omega}_{t},\quad
	\tilde{y}_{t}\sim\mcal{N}(\tilde{y}_{t}|\tilde{H}\tilde{x}_{t},\tilde{E}),\quad
	\tilde{u}_{t}=\tilde{u}(\tilde{y}_{t}),\nonumber\\
	&J[\tilde{u}]:=\lim_{T\to\infty}\frac{1}{T}\mb{E}\left[\int_{0}^{T}\left(\tilde{x}_{t}^{\top}\tilde{Q}\tilde{x}_{t}-2\tilde{x}_{t}^{\top}\tilde{S}\tilde{u}_{t}+\tilde{u}_{t}^{\top}\tilde{R}\tilde{u}_{t}\right)dt\right],\quad
	\tilde{u}=\arg\min_{\tilde{u}}J[\tilde{u}],
\end{align}
where the parameters are given by
\begin{align}
	\tilde{A}:=\left(\begin{array}{cc}
		-1&0\\
		0&0\\
	\end{array}\right),\quad 
	\tilde{B}:=\left(\begin{array}{cc}
		0&0\\
		0&1\\
	\end{array}\right),\quad 
	\tilde{\sigma}:=\left(\begin{array}{cc}
		\sqrt{D}&0\\
		0&\sqrt{F}\\
	\end{array}\right),\quad
	\tilde{H}:=\left(\begin{array}{cc}
		1&0\\
		0&1\\
	\end{array}
	\right),\quad 
	\tilde{E}:=\left(\begin{array}{cc}
		E&0\\
		0&0\\
	\end{array}
	\right),\nonumber\\
	\tilde{Q}:=\left(\begin{array}{cc}
		Q&0\\
		0&0\\
	\end{array}\right),\quad 
	\tilde{S}:=\left(\begin{array}{cc}
		Q&0\\
		0&0\\
	\end{array}\right),\quad 
	\tilde{R}:=\left(\begin{array}{cc}
		Q&0\\
		0&M\\
	\end{array}\right). 
	\label{eq: LQG-parameters}
\end{align}

By employing Pontryagin's minimum principle on the probability density function space \cite{bensoussan_master_2015,bensoussan_interpretation_2017,carmona_probabilistic_2018,carmona_probabilistic_2018-1,tottori_memory-limited_2022,tottori_forward-backward_2023,tottori_decentralized_2023,tottori_resource_2025,tottori_theory_2025}, the solution that satisfies the optimality condition $\delta \tilde{J}/\delta \tilde{u}=0$ is given by 
\begin{align}
	\tilde{u}^{*}(\tilde{y})
	&=-\tilde{R}^{-1}\left(\tilde{\Pi}\tilde{B}-\tilde{S}\right)^{\top}\tilde{K}\tilde{y}, 
	\label{eq: LQG-optimality condition}
\end{align}
where $\tilde{K}:=\tilde{\Sigma}\tilde{H}^{\top}(\tilde{E}+\tilde{H}\tilde{\Sigma}\tilde{H}^{\top})^{-1}$ is the estimation gain matrix because it satisfies $\tilde{K}\tilde{y}=\mb{E}_{p(\tilde{x}|\tilde{y})}[\tilde{x}]$. 
$\tilde{\Sigma}$ and $\tilde{\Pi}$ are the covariance matrix and the control gain matrix, respectively, which are the solutions of the following steady-state equations: 
\begin{align}
	O&=\tilde{\sigma}\tilde{\sigma}^{\top}+\left(\tilde{A}-\tilde{B}\tilde{R}^{-1}\left(\tilde{\Pi}\tilde{B}-\tilde{S}\right)^{\top}\tilde{K}\tilde{H}\right)\tilde{\Sigma}+\tilde{\Sigma}\left(\tilde{A}-\tilde{B}\tilde{R}^{-1}\left(\tilde{\Pi}\tilde{B}-\tilde{S}\right)^{\top}\tilde{K}\tilde{H}\right)^{\top},\label{si-eq: ODE of Si}\\
	O&=\tilde{Q}+\tilde{A}^{\top}\tilde{\Pi}+\tilde{\Pi}\tilde{A}-\left(\tilde{\Pi}\tilde{B}-\tilde{S}\right)\tilde{R}^{-1}\left(\tilde{\Pi}\tilde{B}-\tilde{S}\right)^{\top}+\left(I-\tilde{K}\tilde{H}\right)^{\top}\left(\tilde{\Pi}\tilde{B}-\tilde{S}\right)\tilde{R}^{-1}\left(\tilde{\Pi}\tilde{B}-\tilde{S}\right)^{\top}\left(I-\tilde{K}\tilde{H}\right).\label{si-eq: ODE of Pi}
\end{align}
Therefore, the optimal solution [Eq. (\ref{eq: LQG-optimality condition})] is obtained by solving Eqs. (\ref{si-eq: ODE of Si}) and (\ref{si-eq: ODE of Pi}). 

Eq. (\ref{si-eq: ODE of Pi}) is similar to the Riccati equation in the standard optimal control theory \cite{yong_stochastic_1999,nisio_stochastic_2015,bensoussan_estimation_2018}, but the last term is additional because this work considers partial and noisy observations. 
Eq. (\ref{si-eq: ODE of Pi}) is referred to as the observation-based Riccati equation \cite{tottori_resource_2025,tottori_theory_2025}. 
If the extended observation $\tilde{y}_{t}$ is identical to the extended state $\tilde{x}_{t}$, i.e., $\tilde{H}=I$ and $\tilde{E}=O$, the last term of the observation-based Riccati equation vanishes, reducing it to the standard Riccati equation.

From Eqs. (\ref{eq: LQG-parameters}) and (\ref{eq: LQG-optimality condition}), the optimal state estimator $\hat{x}^{*}$ and memory control $v^{*}$ are calculated as follows: 
\begin{align}
	\left(\begin{array}{c}
		\hat{x}^{*}(y,z)\\
		v^{*}(y,z)\\
	\end{array}\right)
	&=-\left(\begin{array}{cc}
		Q&0\\
		0&M\\
	\end{array}\right)^{-1}\left(\left(\begin{array}{cc}
		\Pi_{xx}&\Pi_{xz}\\
		\Pi_{zx}&\Pi_{zz}\\
	\end{array}\right)\left(\begin{array}{cc}
		0&0\\
		0&1\\
	\end{array}\right)-\left(\begin{array}{cc}
		Q&0\\
		0&0\\
	\end{array}\right)\right)^{\top}
	\left(\begin{array}{cc}
		K_{xy}&K_{xz}\\
		K_{zy}&K_{zz}\\
	\end{array}\right)
	\left(\begin{array}{c}
		y\\
		z\\
	\end{array}\right)\nonumber\\
	&=\left(\begin{array}{c}
		K_{xy}y+K_{xz}z\\
		-M^{-1}\Pi_{zx}\left(K_{xy}y+K_{xz}z\right)-M^{-1}\Pi_{zz}\left(K_{zy}y+K_{zz}z\right)\\
	\end{array}\right). 
\end{align}
The estimation gain matrix $\tilde{K}:=\tilde{\Sigma}\tilde{H}^{\top}(\tilde{E}+\tilde{H}\tilde{\Sigma}\tilde{H}^{\top})^{-1}$ is calculated as follows: 
\begin{align}
	\tilde{K}
	&=\left(\begin{array}{cc}
		\Sigma_{xx}&\Sigma_{xz}\\
		\Sigma_{zx}&\Sigma_{zz}\\
	\end{array}\right)\left(\begin{array}{cc}
		1&0\\
		0&1\\
	\end{array}\right)^{\top}\left(\left(\begin{array}{cc}
		E&0\\
		0&0\\
	\end{array}\right)+\left(\begin{array}{cc}
		1&0\\
		0&1\\
	\end{array}\right)\left(\begin{array}{cc}
		\Sigma_{xx}&\Sigma_{xz}\\
		\Sigma_{zx}&\Sigma_{zz}\\
	\end{array}\right)\left(\begin{array}{cc}
		1&0\\
		0&1\\
	\end{array}\right)^{\top}\right)^{-1}\nonumber\\
	&=\left(\begin{array}{cc}
		\left(E+\Sigma_{x|z}\right)^{-1}\Sigma_{x|z}&\left(E+\Sigma_{x|z}\right)^{-1}E\Sigma_{xz}\Sigma_{zz}^{-1}\\
		0&1\\
	\end{array}\right),
\end{align}
where $\Sigma_{x|z}:=\Sigma_{xx}-\Sigma_{xz}^{2}\Sigma_{zz}^{-1}$ is the conditional variance of the state $x$ given the memory $z$. 
Therefore, the optimal state estimator $\hat{x}^{*}$ and memory control $v^{*}$ are given by 
\begin{align}
	\hat{x}^{*}(y,z)&=\left(E+\Sigma_{x|z}\right)^{-1}\Sigma_{x|z}y+\left(E+\Sigma_{x|z}\right)^{-1}E\Sigma_{xz}\Sigma_{zz}^{-1}z,\\
	v^{*}(y,z)&=-M^{-1}\Pi_{zx}\hat{x}^{*}(y,z)-M^{-1}\Pi_{zz}z.\label{si-eq: optimal control}
\end{align}

To obtain $\hat{x}^{*}(y,z)$ and $v^{*}(y,z)$, it is necessary to solve Eqs. (\ref{si-eq: ODE of Si}) and (\ref{si-eq: ODE of Pi}) to determine $\tilde{\Sigma}$ and $\tilde{\Pi}$. 
However, these equations are analytically intractable due to their multivariable and nonlinear nature. 
Therefore, the previous work investigated them numerically \cite{tottori_resource_2025,tottori_theory_2025}. 
In numerical calculations, the steady-state equation of the precision matrix $\tilde{\Lambda}:=\tilde{\Sigma}^{-1}$, i.e., 
\begin{align}
	O&=-\left(\tilde{A}-\tilde{B}\tilde{R}^{-1}\left(\tilde{\Pi}\tilde{B}-\tilde{S}\right)^{\top}\tilde{K}\tilde{H}\right)^{\top}\tilde{\Lambda}
	-\tilde{\Lambda}\left(\tilde{A}-\tilde{B}\tilde{R}^{-1}\left(\tilde{\Pi}\tilde{B}-\tilde{S}\right)^{\top}\tilde{K}\tilde{H}\right)
	-\tilde{\Lambda}\tilde{\sigma}\tilde{\sigma}^{\top}\tilde{\Lambda},
	\label{si-eq: ODE of La}
\end{align}
is more practical than that of the covariance matrix $\tilde{\Sigma}$ [Eq. (\ref{si-eq: ODE of Si})] because the elements of $\tilde{\Lambda}$ converge to 0 when those of $\tilde{\Sigma}$ diverge to $\infty$. 
Therefore, Eq. (\ref{si-eq: ODE of La}) is employed in numerical experiments instead of Eq. (\ref{si-eq: ODE of Si}). 

\section{Reformulation of Control Function}
To enable analytical treatment, we reformulate the control function as a linear function of the observation $y$ and memory $z$: 
\begin{align}
	v_{\Phi}(y,z):=-\Phi_{zy}y-\Phi_{zz}z, 
	\label{si-eq: linear control function}
\end{align}
where $\Phi_{zy}\in\mb{R}$ and $\Phi_{zz}\geq0$. 
This reformulation does not degrade performance because the optimal control function $v^{*}(y,z)$ [Eq. (\ref{si-eq: optimal control})] is given by the following linear function: 
\begin{align}
	v^{*}(y,z)
	&=-M^{-1}\Pi_{zx}\hat{x}^{*}(y,z)-M^{-1}\Pi_{zz}z\nonumber\\
	&=-M^{-1}\Pi_{zx}\left\{\left(E+\Sigma_{x|z}\right)^{-1}\Sigma_{x|z}y+\left(E+\Sigma_{x|z}\right)^{-1}E\Sigma_{xz}\Sigma_{zz}^{-1}z\right\}-M^{-1}\Pi_{zz}z\nonumber\\
	&=-\underbrace{M^{-1}\Pi_{zx}\left(E+\Sigma_{x|z}\right)^{-1}\Sigma_{x|z}}_{\Phi_{zy}^{*}}y-\underbrace{M^{-1}\left\{\Pi_{zx}\left(E+\Sigma_{x|z}\right)^{-1}E\Sigma_{xz}\Sigma_{zz}^{-1}+\Pi_{zz}\right\}}_{\Phi_{zz}^{*}}z,
\end{align}
where $\Phi_{zy}^{*}$ and $\Phi_{zz}^{*}$ are the control gains that satisfy the optimality condition in the previous work \cite{tottori_resource_2025,tottori_theory_2025}. 

As shown below, this reformulation enables us to derive the analytical expressions of the covariance matrix $\tilde{\Sigma}$ (Sec. \ref{si-subsec: rcg-dcm}) and the objective function $J$ (Sec. \ref{si-subsec: rcg-dof}). 

\subsection{Analytical Expression of Covariance Matrix $\tilde{\Sigma}$}\label{si-subsec: rcg-dcm}
When the control function is given by Eq. (\ref{si-eq: linear control function}), the steady-state equation for the covariance matrix $\tilde{\Sigma}$ becomes
\begin{align}
	O&=\tilde{D}+\left(\tilde{A}-\tilde{B}\tilde{\Phi}\tilde{H}\right)\tilde{\Sigma}+\tilde{\Sigma}\left(\tilde{A}-\tilde{B}\tilde{\Phi}\tilde{H}\right)^{\top}. \label{si-eq: ODE of Si ver2}
\end{align}
If the control gain matrix is given by $\tilde{\Phi}=\tilde{R}^{-1}(\tilde{\Pi}\tilde{B}-\tilde{S})^{\top}\tilde{K}$, Eq. (\ref{si-eq: ODE of Si ver2}) becomes Eq. (\ref{si-eq: ODE of Si}). 
However, it is analytically intractable because $\tilde{K}:=\tilde{\Sigma}\tilde{H}^{\top}(\tilde{E}+\tilde{H}\tilde{\Sigma}\tilde{H}^{\top})^{-1}$ introduces nonlinearity. 
In contrast, when $\tilde{\Phi}$ is fixed independently of $\tilde{\Sigma}$, this nonlinearity is avoided, and Eq. (\ref{si-eq: ODE of Si ver2}) becomes linear, allowing for analytical solutions. 

Substituting the parameters [Eq. (\ref{eq: LQG-parameters})] into Eq. (\ref{si-eq: ODE of Si ver2}), we obtain 
\begin{align}
	\left(\begin{array}{cc}
		0&0\\
		0&0\\
	\end{array}\right)
	&=\left(\begin{array}{cc}
		D&0\\
		0&F\\
	\end{array}\right)
	+\left(\left(\begin{array}{cc}
		-1&0\\
		0&0\\
	\end{array}\right)
	-\left(\begin{array}{cc}
		0&0\\
		0&1\\
	\end{array}\right)\left(\begin{array}{cc}
		\Phi_{yy}&\Phi_{yz}\\
		\Phi_{zy}&\Phi_{zz}\\
	\end{array}\right)\left(\begin{array}{cc}
		1&0\\
		0&1\\
	\end{array}\right)\right)\left(\begin{array}{cc}
		\Sigma_{xx}&\Sigma_{xz}\\
		\Sigma_{zx}&\Sigma_{zz}\\
	\end{array}\right)\nonumber\\
	&\quad+\left(\begin{array}{cc}
		\Sigma_{xx}&\Sigma_{xz}\\
		\Sigma_{zx}&\Sigma_{zz}\\
	\end{array}\right)\left(\left(\begin{array}{cc}
		-1&0\\
		0&0\\
	\end{array}\right)
	-\left(\begin{array}{cc}
		0&0\\
		0&1\\
	\end{array}\right)\left(\begin{array}{cc}
		\Phi_{yy}&\Phi_{yz}\\
		\Phi_{zy}&\Phi_{zz}\\
	\end{array}\right)\left(\begin{array}{cc}
		1&0\\
		0&1\\
	\end{array}\right)\right)^{\top}\nonumber\\
	&=\left(\begin{array}{cc}
		D&0\\
		0&F\\
	\end{array}\right)
	+\left(\begin{array}{cc}
		-1&0\\
		-\Phi_{zy}&-\Phi_{zz}\\
	\end{array}\right)\left(\begin{array}{cc}
		\Sigma_{xx}&\Sigma_{xz}\\
		\Sigma_{zx}&\Sigma_{zz}\\
	\end{array}\right)+\left(\begin{array}{cc}
		\Sigma_{xx}&\Sigma_{xz}\\
		\Sigma_{zx}&\Sigma_{zz}\\
	\end{array}\right)\left(\begin{array}{cc}
		-1&-\Phi_{zy}\\
		0&-\Phi_{zz}\\
	\end{array}\right).
\end{align}
As a result, $\Sigma_{xx}$, $\Sigma_{xz}$, and $\Sigma_{zz}$ are given by 
\begin{align}
	\Sigma_{xx}&=\frac{D}{2},\label{si-eq: Sigma-xx}\\
	\Sigma_{xz}&=-\frac{\Phi_{zy}\Sigma_{xx}}{\Phi_{zz}+1}=-\frac{\Phi_{zy}D}{2\left(\Phi_{zz}+1\right)},\label{si-eq: Sigma-xz}\\
	\Sigma_{zz}&=\frac{F-2\Phi_{zy}\Sigma_{xz}}{2\Phi_{zz}}=\frac{\Phi_{zy}^{2}D+\left(\Phi_{zz}+1\right)F}{2\left(\Phi_{zz}+1\right)\Phi_{zz}}\label{si-eq: Sigma-zz}.
\end{align}
We note that $\Sigma_{xz}=\Sigma_{zx}$ holds because $\tilde{\Sigma}$ is a symmetric matrix. 
Furthermore, the conditional variance $\Sigma_{x|z}:=\Sigma_{xx}-\Sigma_{xz}^{2}\Sigma_{zz}^{-1}$ is given by 
\begin{align}
	\Sigma_{x|z}
	&=\frac{\Phi_{zy}^{2}D+\left(\Phi_{zz}+1\right)^{2}F}{\left(\Phi_{zz}+1\right)\left\{\Phi_{zy}^{2}D+\left(\Phi_{zz}+1\right)F\right\}}\frac{D}{2}.\label{si-eq: Sigma-x-z}
\end{align}

\subsection{Analytical Expression of Objective Function $J$}\label{si-subsec: rcg-dof}
The objective function is given by 
\begin{align}
	J:=\lim_{T\to\infty}\frac{1}{T}\mb{E}\left[\int_{0}^{T}\left(\tilde{x}_{t}^{\top}\tilde{Q}\tilde{x}_{t}-2\tilde{x}_{t}^{\top}\tilde{S}\tilde{u}_{t}+\tilde{u}_{t}^{\top}\tilde{R}\tilde{u}_{t}\right)dt\right].\label{si-eq: OF-ver1}
\end{align}
When the extended control is given by $\tilde{u}_{t}:=-\tilde{\Phi}\tilde{y}_{t}$, it becomes
\begin{align}
	J=\lim_{T\to\infty}\frac{1}{T}\mb{E}\left[\int_{0}^{T}\left(\tilde{x}_{t}^{\top}\tilde{Q}\tilde{x}_{t}+2\tilde{x}_{t}^{\top}\tilde{S}\tilde{\Phi}\tilde{y}_{t}+\tilde{y}_{t}^{\top}\tilde{\Phi}^{\top}\tilde{R}\tilde{\Phi}\tilde{y}_{t}\right)dt\right].\label{si-eq: OF-ver2}
\end{align}
In the long-time limit $T\to\infty$, the probability distribution of the extended state $\tilde{x}$ converges to the following stationary distribution: 
\begin{align}
	p(\tilde{x})=\mcal{N}\left(\tilde{x}\left|0,\tilde{\Sigma}\right.\right).\label{si-eq: OF-PD-ver1}
\end{align}
For simplicity, we consider the case where the mean of $\tilde{x}$ is zero, whereas the extension to nonzero means is straightforward \cite{tottori_theory_2025}. 
From Eq. (\ref{si-eq: OF-ver2}) and (\ref{si-eq: OF-PD-ver1}), we obtain 
\begin{align}
	J&=\int \int p(\tilde{y}|\tilde{x})p(\tilde{x})\left[\tilde{x}^{\top}\tilde{Q}\tilde{x}+2\tilde{x}^{\top}\tilde{S}\tilde{\Phi}\tilde{y}+\tilde{y}^{\top}\tilde{\Phi}^{\top}\tilde{R}\tilde{\Phi}\tilde{y}\right]d\tilde{x}d\tilde{y}\nonumber\\
	&={\rm tr}\left\{\left(\tilde{Q}+2\tilde{S}\tilde{\Phi}\tilde{H}+\tilde{H}^{\top}\tilde{\Phi}^{\top}\tilde{R}\tilde{\Phi}\tilde{H}\right)\tilde{\Sigma}+\tilde{\Phi}^{\top}\tilde{R}\tilde{\Phi}\tilde{E}\right\}.\label{si-eq: OF-ver3}
\end{align}
The parameters $\tilde{Q}$, $\tilde{S}$, $\tilde{R}$, $\tilde{H}$, and $\tilde{E}$ are given by Eq. (\ref{eq: LQG-parameters}). 
In addition, from 
\begin{align}
	\tilde{u}:=-\tilde{\Phi}\tilde{y}
	\quad\Rightarrow\quad
	\left(\begin{array}{c}
		\hat{x}^{*}\\
		v_{\Phi}\\
	\end{array}\right)&=
	-\left(\begin{array}{cc}
		-\left(E+\Sigma_{x|z}\right)^{-1}\Sigma_{x|z}&-\left(E+\Sigma_{x|z}\right)^{-1}E\Sigma_{xz}\Sigma_{zz}^{-1}\\
		\Phi_{zy}&\Phi_{zz}\\
	\end{array}\right)\left(\begin{array}{c}
		y\\
		z\\
	\end{array}\right),
\end{align}
$\tilde{\Phi}$ is given by 
\begin{align}
	\left(\begin{array}{cc}
		\Phi_{xx}&\Phi_{xz}\\
		\Phi_{zx}&\Phi_{zz}\\
	\end{array}\right)&=\left(\begin{array}{cc}
		-\left(E+\Sigma_{x|z}\right)^{-1}\Sigma_{x|z}&-\left(E+\Sigma_{x|z}\right)^{-1}E\Sigma_{xz}\Sigma_{zz}^{-1}\\
		\Phi_{zx}&\Phi_{zz}\\
	\end{array}\right). 
\end{align}
As a result, Eq. (\ref{si-eq: OF-ver3}) is calculated as follows: 
\begin{align}
	J=Q\frac{E\Sigma_{x|z}}{E+\Sigma_{x|z}}+M\left\{\Phi_{zy}^{2}\left(\Sigma_{xx}+E\right)+2\Phi_{zy}\Phi_{zz}\Sigma_{xz}+\Phi_{zz}^{2}\Sigma_{zz}\right\}.
\end{align}
The first and second terms correspond to the state estimation error and the memory control cost, respectively. 
To clarify this point, we decompose the objective function $J$ into the state estimation error $J_{Q}$ and the memory control cost $J_{M}$ as follows: 
\begin{align}
	&J=J_{Q}+J_{M},\label{si-eq: J=JQ+JM}\\
	&J_{Q}:=Q\frac{E\Sigma_{x|z}}{E+\Sigma_{x|z}},\\
	&J_{M}:=M\left\{\Phi_{zy}^{2}\left(\Sigma_{xx}+E\right)+2\Phi_{zy}\Phi_{zz}\Sigma_{xz}+\Phi_{zz}^{2}\Sigma_{zz}\right\}.
\end{align}
Substituting the analytical expressions of $\Sigma_{xx}$, $\Sigma_{xz}$, $\Sigma_{zz}$, and $\Sigma_{x|z}$ [Eqs. (\ref{si-eq: Sigma-xx})--(\ref{si-eq: Sigma-x-z})], $J_{Q}$ and $J_{M}$ can be rewritten as follows: 
\begin{align}
	&J_{Q}=Q\frac{\left\{\Phi_{zy}^{2}D+\left(\Phi_{zz}+1\right)^{2}F\right\}DE}{\Phi_{zy}^{2}\left\{D+2\left(\Phi_{zz}+1\right)E\right\}D+\left(\Phi_{zz}+1\right)^{2}\left(D+2E\right)F},\label{si-eq: JQ}\\
	&J_{M}=M\frac{\Phi_{zy}^{2}D+2\Phi_{zy}^{2}\left(\Phi_{zz}+1\right)E+\Phi_{zz}\left(\Phi_{zz}+1\right)F}{2\left(\Phi_{zz}+1\right)}.\label{si-eq: JM}
\end{align}

\section{Optimization of Only $\Phi_{zy}$ at Fixed $\Phi_{zz}$}
We first analyze the stationary condition $\partial J/\partial \Phi_{zy}=0$ at fixed $\Phi_{zz}$. 
From Eqs. (\ref{si-eq: J=JQ+JM}), (\ref{si-eq: JQ}), and (\ref{si-eq: JM}), the stationary condition $\partial J/\partial \Phi_{zy}=0$ is given by 
\begin{align}
	&0=\frac{\partial J}{\partial\Phi_{zy}}=\frac{\partial J_{Q}}{\partial\Phi_{zy}}+\frac{\partial J_{M}}{\partial\Phi_{zy}},\label{si-eq: pJpPhizy-ver1}\\
	&\frac{\partial J_{Q}}{\partial\Phi_{zy}}=Q\frac{-4\Phi_{zy}\Phi_{zz}\left(\Phi_{zz}+1\right)^{2}D^{2}E^{2}F}{\left[\Phi_{zy}^{2}\left\{D+2\left(\Phi_{zz}+1\right)E\right\}D+\left(\Phi_{zz}+1\right)^{2}\left(D+2E\right)F\right]^{2}},\label{si-eq: pJpPhizy-ver1-Q}\\
	&\frac{\partial J_{M}}{\partial\Phi_{zy}}=M\frac{\Phi_{zy}\left\{D+2\left(\Phi_{zz}+1\right)E\right\}}{\Phi_{zz}+1}.\label{si-eq: pJpPhizy-ver1-M}
\end{align}
Eq. (\ref{si-eq: pJpPhizy-ver1}) is calculated as follows: 
\begin{align}
	&0=\frac{\partial J}{\partial\Phi_{zy}}=\frac{\partial J_{Q}}{\partial\Phi_{zy}}+\frac{\partial J_{M}}{\partial\Phi_{zy}}\nonumber\\
	\Rightarrow\ 
	&0=Q\frac{-4\Phi_{zy}\Phi_{zz}\left(\Phi_{zz}+1\right)^{2}D^{2}E^{2}F}{\left[\Phi_{zy}^{2}\left\{D+2\left(\Phi_{zz}+1\right)E\right\}D+\left(\Phi_{zz}+1\right)^{2}\left(D+2E\right)F\right]^{2}}
	+M\frac{\Phi_{zy}\left\{D+2\left(\Phi_{zz}+1\right)E\right\}}{\Phi_{zz}+1}\nonumber\\
	\Rightarrow\ 
	&0=\Phi_{zy}\left[-4Q\Phi_{zz}\left(\Phi_{zz}+1\right)^{3}D^{2}E^{2}F\right.\nonumber\\
	&\ \ \ \left.+M\left\{D+2\left(\Phi_{zz}+1\right)E\right\}\left[\left\{D+2\left(\Phi_{zz}+1\right)E\right\}D\Phi_{zy}^{2}+\left(\Phi_{zz}+1\right)^{2}\left(D+2E\right)F\right]^{2}\right]\nonumber\\
	\Rightarrow\ 
	&0=\Phi_{zy}\left[-4Q\Phi_{zz}\left(\Phi_{zz}+1\right)^{3}D^{2}E^{2}F\right.\nonumber\\
	&\ \ \ \left.+M\left\{D+2\left(\Phi_{zz}+1\right)E\right\}^{3}D^{2}\left(\Phi_{zy}^{2}+\frac{\left(\Phi_{zz}+1\right)^{2}\left(D+2E\right)F}{\left\{D+2\left(\Phi_{zz}+1\right)E\right\}D}\right)^{2}\right]\nonumber\\
	\Rightarrow\ 
	&0=\Phi_{zy}\left[-4\frac{Q\Phi_{zz}\left(\Phi_{zz}+1\right)^{3}E^{2}F}{M\left\{D+2\left(\Phi_{zz}+1\right)E\right\}^{3}}
	+\left(\Phi_{zy}^{2}+\frac{\left(\Phi_{zz}+1\right)^{2}\left(D+2E\right)F}{\left\{D+2\left(\Phi_{zz}+1\right)E\right\}D}\right)^{2}\right]\nonumber\\
	\Rightarrow\ 
	&0=\Phi_{zy}\underbrace{\left(\Phi_{zy}^{2}+\frac{\left(\Phi_{zz}+1\right)^{2}\left(D+2E\right)F}{\left\{D+2\left(\Phi_{zz}+1\right)E\right\}D}+2\sqrt{\frac{Q\Phi_{zz}\left(\Phi_{zz}+1\right)^{3}E^{2}F}{M\left\{D+2\left(\Phi_{zz}+1\right)E\right\}^{3}}}\right)}_{\neq0}\nonumber\\
	&\ \ \ \times\left(\Phi_{zy}^{2}+\frac{\left(\Phi_{zz}+1\right)^{2}\left(D+2E\right)F}{\left\{D+2\left(\Phi_{zz}+1\right)E\right\}D}-2\sqrt{\frac{Q\Phi_{zz}\left(\Phi_{zz}+1\right)^{3}E^{2}F}{M\left\{D+2\left(\Phi_{zz}+1\right)E\right\}^{3}}}\right)\nonumber\\
	\Rightarrow\ 
	&0=\Phi_{zy}\left(\Phi_{zy}^{2}+\frac{\left(\Phi_{zz}+1\right)^{2}\left(D+2E\right)F}{\left\{D+2\left(\Phi_{zz}+1\right)E\right\}D}-2\sqrt{\frac{Q\Phi_{zz}\left(\Phi_{zz}+1\right)^{3}E^{2}F}{M\left\{D+2\left(\Phi_{zz}+1\right)E\right\}^{3}}}\right).
\end{align}
As a result, the solutions that satisfy $\partial J/\partial \Phi_{zy}=0$ are given by $\Phi_{zy}^{*}=0$ or 
\begin{align}
	\Phi_{zy}^{*}
	&=\pm\sqrt{2\sqrt{\frac{Q\Phi_{zz}\left(\Phi_{zz}+1\right)^{3}E^{2}F}{M\left\{D+2\left(\Phi_{zz}+1\right)E\right\}^{3}}}-\frac{\left(\Phi_{zz}+1\right)^{2}\left(D+2E\right)F}{\left\{D+2\left(\Phi_{zz}+1\right)E\right\}D}}\nonumber\\
	&=\pm\sqrt{\frac{\left(\Phi_{zz}+1\right)^{2}\left(D+2E\right)F}{\left\{D+2\left(\Phi_{zz}+1\right)E\right\}D}\left(\sqrt{\Theta}-1\right)},\label{si-eq: Phizy-star-ver1}
\end{align}
where 
\begin{align}
	\Theta
	&:=\frac{4\Phi_{zz}D^{2}E^{2}}{\left(\Phi_{zz}+1\right)\left\{D+2\left(\Phi_{zz}+1\right)E\right\}\left(D+2E\right)^{2}}\frac{Q}{MF}.\label{si-eq: Theta-ver1}
\end{align}
As discussed in the main text, these solutions account for the scaling relation $Q/MF$ and the nonmonotonic phase transitions with respect to $D$ and $E$. 
However, they fail to capture the discontinuity of the resource-induced phase transitions. 
This means that the stationary condition $\partial J/\partial \Phi_{zy}=0$ alone cannot explain the emergence of the discontinuity. 

\section{Simultaneous Optimization of $\Phi_{zy}$ and $\Phi_{zz}$}
To account for the discontinuity of the resource-induced phase transitions, we analyze $\partial J/\partial \Phi_{zz}=0$ as well as $\partial J/\partial \Phi_{zy}=0$. 
From Eqs. (\ref{si-eq: J=JQ+JM}), (\ref{si-eq: JQ}), and (\ref{si-eq: JM}), $\partial J/\partial \Phi_{zz}=0$ is given by 
\begin{align}
	&0=\frac{\partial J}{\partial\Phi_{zz}}=\frac{\partial J_{Q}}{\partial\Phi_{zz}}+\frac{\partial J_{M}}{\partial\Phi_{zz}},\label{si-eq: pJpPhizz-ver1}\\
	&\frac{\partial J_{Q}}{\partial\Phi_{zz}}=Q\frac{2\Phi_{zy}^{2}D^{2}E^{2}\left\{\left(\Phi_{zz}+1\right)\left(\Phi_{zz}-1\right)F-\Phi_{zy}^{2}D\right\}}{\left[\Phi_{zy}^{2}\left\{D+2\left(\Phi_{zz}+1\right)E\right\}D+\left(\Phi_{zz}+1\right)^{2}\left(D+2E\right)F\right]^{2}},\label{si-eq: pJpPhizz-ver1-Q}\\
	&\frac{\partial J_{M}}{\partial\Phi_{zz}}=M\frac{\left(\Phi_{zz}+1\right)^{2}F-\Phi_{zy}^{2}D}{2\left(\Phi_{zz}+1\right)^{2}}.\label{si-eq: pJpPhizz-ver1-M}
\end{align}
This stationary condition is analytically intractable because it is a high-order equation of $\Phi_{zz}$ and $\Phi_{zy}$. 
Nevertheless, we are able to derive some analytical results for this stationary condition. 

This section is organized as follows: 
In Sec. \ref{si-subsec: Phizz-scaling}, we show that the scaling relation $Q/MF$ still holds in the simultaneous optimization of $\Phi_{zy}$ and $\Phi_{zz}$ without approximation. 
In Sec. \ref{si-subsec: Phizz-decomposition}, we show that the solutions to $\partial J_{Q}/\partial \Phi_{zz}=0$ and $\partial J_{M}/\partial \Phi_{zz}=0$ provide upper and lower bounds, respectively, for the solution to $\partial J/\partial \Phi_{zz}=0$. 
In Sec. \ref{si-subsec: Phizz-Q} and \ref{si-subsec: Phizz-M}, we analyze the conditions under which $\partial J_{Q}/\partial \Phi_{zz}=0$ and $\partial J_{M}/\partial \Phi_{zz}=0$ intersect with $\partial J/\partial \Phi_{zy}=0$, respectively. 
The former gives a necessary condition under which $\partial J/\partial \Phi_{zz}=0$ intersect with $\partial J/\partial \Phi_{zy}=0$, whereas the latter gives a sufficient condition. 

\subsection{Scaling Relation}\label{si-subsec: Phizz-scaling}
In this subsection, we analytically demonstrate that the scaling relation $Q/MF$ remains valid even in the simultaneous optimization of $\Phi_{zy}$ and $\Phi_{zz}$. 
From Eqs. (\ref{si-eq: pJpPhizz-ver1}), (\ref{si-eq: pJpPhizz-ver1-Q}), and (\ref{si-eq: pJpPhizz-ver1-M}), the stationary condition $\partial J/\partial \Phi_{zz}=0$ is calculated as follows: 
\begin{align}
	0=\frac{QD}{MF}\frac{2\left(\frac{D}{F}\Phi_{zy}^{2}\right)\left(\frac{E}{D}\right)^{2}\left\{\left(\Phi_{zz}+1\right)\left(\Phi_{zz}-1\right)-\left(\frac{D}{F}\Phi_{zy}^{2}\right)\right\}}{\left[\left\{1+2\left(\Phi_{zz}+1\right)\frac{E}{D}\right\}\left(\frac{D}{F}\Phi_{zy}^{2}\right)+\left(\Phi_{zz}+1\right)^{2}\left(1+2\frac{E}{D}\right)\right]^{2}}+\frac{\left(\Phi_{zz}+1\right)^{2}-\left(\frac{D}{F}\Phi_{zy}^{2}\right)}{2\left(\Phi_{zz}+1\right)^{2}}.
	\label{si-eq: pJpPhizz-ver2}
\end{align}
By introducing the reparameterization
\begin{align}
	\alpha:=\frac{Q}{M},\quad
	\beta:=\frac{E}{D},\quad
	\gamma:=\frac{F}{D},\quad
	\delta:=\alpha\gamma^{-1},
	\label{si-eq: reparameters}
\end{align}
Eq. (\ref{si-eq: pJpPhizz-ver2}) is rewritten as 
\begin{align}
	0=\delta\frac{2\left(\gamma^{-1}\Phi_{zy}^{2}\right)\beta^{2}\left\{\left(\Phi_{zz}+1\right)\left(\Phi_{zz}-1\right)-\left(\gamma^{-1}\Phi_{zy}^{2}\right)\right\}}{\left[\left\{1+2\left(\Phi_{zz}+1\right)\beta\right\}\left(\gamma^{-1}\Phi_{zy}^{2}\right)+\left(\Phi_{zz}+1\right)^{2}\left(1+2\beta\right)\right]^{2}}+\frac{\left(\Phi_{zz}+1\right)^{2}-\left(\gamma^{-1}\Phi_{zy}^{2}\right)}{2\left(\Phi_{zz}+1\right)^{2}}.
	\label{si-eq: pJpPhizz-ver3}
\end{align}
This result shows that the functional form of $\Phi_{zy}^{2}$ is given by 
\begin{align}
	\Phi_{zy}^{2}=\gamma f\left(\Phi_{zz},\beta,\delta\right),\label{si-eq: pJpPhizz-ver4}
\end{align}
where $f$ is a function of $\Phi_{zz}$, $\beta$, and $\delta$ that satisfies Eq. (\ref{si-eq: pJpPhizz-ver3}). 
By using Eq. (\ref{si-eq: reparameters}), the solution to the stationary condition $\partial J/\partial \Phi_{zy}=0$ [Eqs. (\ref{si-eq: Phizy-star-ver1}) and (\ref{si-eq: Theta-ver1})] is rewritten as follows: 
\begin{align}
	\Phi_{zy}^{2}=\gamma\frac{(\Phi_{zz}+1)^{2}(1+2\beta)}{\left\{1+2(\Phi_{zz}+1)\beta\right\}}\left(\sqrt{\frac{4\Phi_{zz}\beta^{2}}{(\Phi_{zz}+1)\left\{1+2(\Phi_{zz}+1)\beta\right\}(1+2\beta)^{2}}\delta}-1\right).\label{si-eq: Phizy-star-ver2}
\end{align}
Substituting Eq. (\ref{si-eq: pJpPhizz-ver4}) into Eq. (\ref{si-eq: Phizy-star-ver2}), we obtain
\begin{align}
	&f\left(\Phi_{zz},\beta,\delta\right)
	=\frac{(\Phi_{zz}+1)^{2}(1+2\beta)}{\left\{1+2(\Phi_{zz}+1)\beta\right\}}\left(\sqrt{\frac{4\Phi_{zz}\beta^{2}}{(\Phi_{zz}+1)\left\{1+2(\Phi_{zz}+1)\beta\right\}(1+2\beta)^{2}}\delta}-1\right).\label{si-eq: pJpPhizz-ver5}
\end{align}
As a result, the functional form of $\Phi_{zz}$ is given by 
\begin{align}
	\Phi_{zz}=g\left(\beta,\delta\right),\label{si-eq: pJpPhizz-ver6}
\end{align}
where $g$ is a function of $\beta$ and $\delta$ that satisfies Eq. (\ref{si-eq: pJpPhizz-ver5}). 
Furthermore, the functional form of $\Phi_{zy}^{2}$ is given by 
\begin{align}
	\Phi_{zy}^{2}=\gamma h\left(\beta,\delta\right),\label{si-eq: pJpPhizz-ver7}
\end{align}
where $h\left(\beta,\delta\right):=f\left(g\left(\beta,\delta\right),\beta,\delta\right)$. 
Eqs. (\ref{si-eq: pJpPhizz-ver6}) and (\ref{si-eq: pJpPhizz-ver7}) show that the phase transition depends only on $\beta$ and $\delta$. 
$\gamma$ determines only the scale of $\Phi_{zy}^{2}$ and does not affect the phase transition except through $\delta:=\alpha\gamma^{-1}$. 
In particular, since $\delta$ includes the scaling relation $Q/MF$, it holds even in the simultaneous optimization of $\Phi_{zy}$ and $\Phi_{zz}$. 

\begin{figure*}
	\includegraphics[width=170mm]{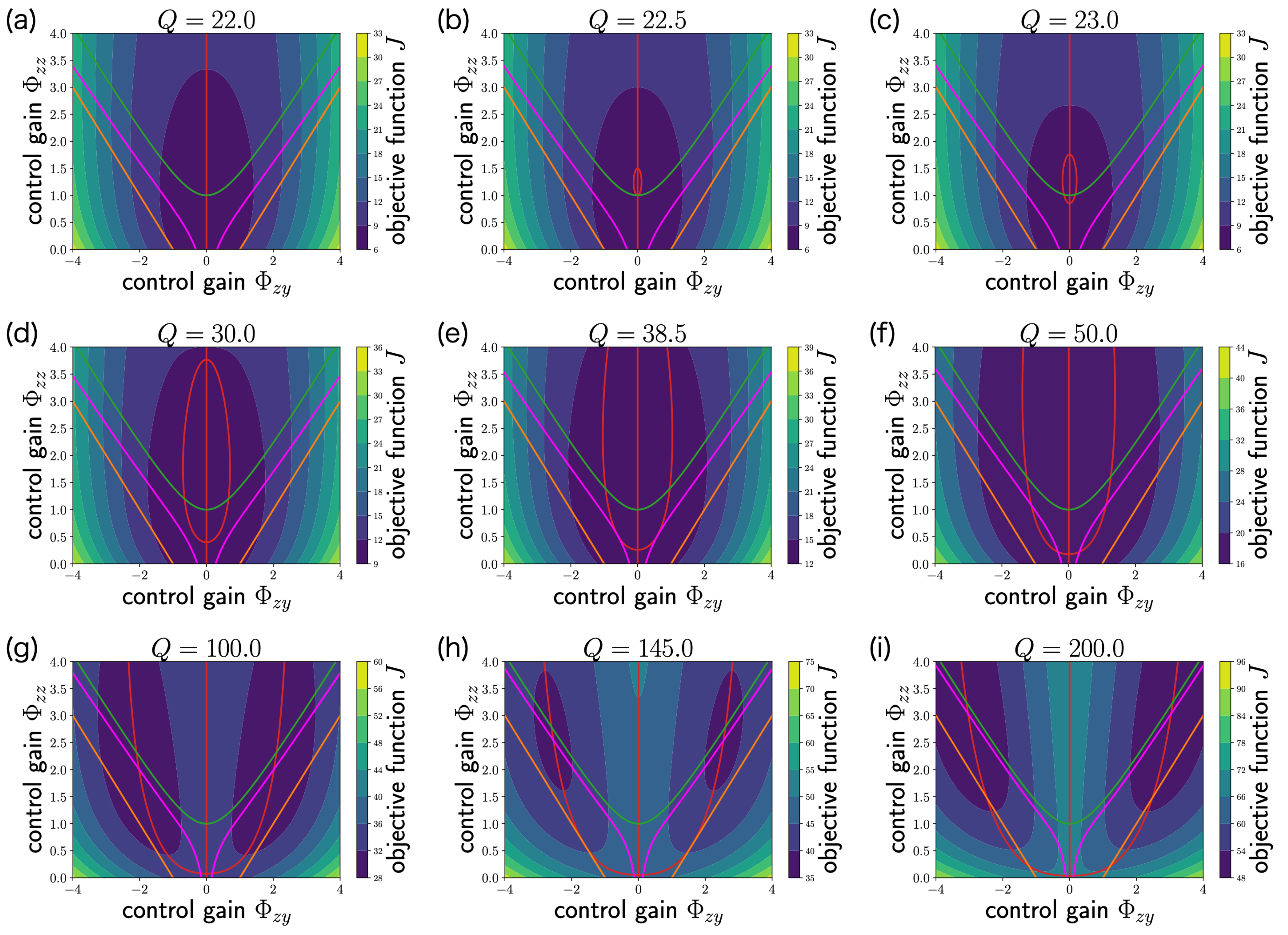}
	\caption{\label{fig-S1}
	$J$ as functions of $\Phi_{zy}$ and $\Phi_{zz}$. 
	Red, magenta, green, and orange curves indicate $\partial J/\partial \Phi_{zty}=0$, $\partial J/\partial \Phi_{zz}=0$, $\partial J_{Q}/\partial \Phi_{zz}=0$, and $\partial J_{M}/\partial \Phi_{zz}=0$, respectively. 
	The rest of the parameters are set to 1. 
	}
\end{figure*}
\subsection{Relationship between $\partial J/\partial \Phi_{zz}=0$, $\partial J_{Q}/\partial \Phi_{zz}=0$, and $\partial J_{M}/\partial \Phi_{zz}=0$}\label{si-subsec: Phizz-decomposition}
Although $\partial J/\partial \Phi_{zz}=0$ is analytically intractable, $\partial J_{Q}/\partial \Phi_{zz}=0$ and $\partial J_{M}/\partial \Phi_{zz}=0$ are more tractable. 
Specifically, from Eqs. (\ref{si-eq: pJpPhizz-ver1-Q}) and (\ref{si-eq: pJpPhizz-ver1-M}), we obtain the following solutions: 
\begin{align}
	&\frac{\partial J_{Q}}{\partial\Phi_{zz}}=0
	\ \Rightarrow\ 
	\gamma^{-1}\Phi_{zy}^{2}=\left(\Phi_{zz}+1\right)\left(\Phi_{zz}-1\right)
	\ \Rightarrow\ 
	\Phi_{zz}=\sqrt{\gamma^{-1}\Phi_{zy}^{2}+1},\label{si-eq: pJpPhizz-ver2-Q}\\
	&\frac{\partial J_{M}}{\partial\Phi_{zz}}=0
	\ \Rightarrow\ 
	\gamma^{-1}\Phi_{zy}^{2}=\left(\Phi_{zz}+1\right)^{2}
	\ \Rightarrow\ 
	\Phi_{zz}=\sqrt{\gamma^{-1}\Phi_{zy}^{2}}-1, \label{si-eq: pJpPhizz-ver2-M}
\end{align}
where $\gamma:=F/D$. 
In this subsection, we demonstrate that the values of $\Phi_{zz}$ that satisfy $\partial J_{Q}/\partial \Phi_{zz}=0$ [Fig. \ref{fig-S1}, green] and $\partial J_{M}/\partial \Phi_{zz}=0$ [Fig. \ref{fig-S1}, orange] serve as upper and lower bounds, respectively, for the value of $\Phi_{zz}$ satisfying $\partial J/\partial \Phi_{zz}=0$ [Fig. \ref{fig-S1}, magenta]. 
This result enables us to derive the necessary and sufficient conditions under which $\partial J/\partial \Phi_{zy}=0$ and $\partial J/\partial \Phi_{zz}=0$ intersect. 
The details are discussed in Sec. \ref{si-subsec: Phizz-Q} and \ref{si-subsec: Phizz-M}. 

From Eq. (\ref{si-eq: pJpPhizz-ver3}), $\partial J/\partial \Phi_{zz}=0$ becomes
\begin{align}
	w\left(\gamma^{-1}\Phi_{zy}^{2},\Phi_{zz},\beta,\delta\right)\left\{\left(\Phi_{zz}+1\right)\left(\Phi_{zz}-1\right)-\gamma^{-1}\Phi_{zy}^{2}\right\}+\left(\Phi_{zz}+1\right)^{2}-\gamma^{-1}\Phi_{zy}^{2}=0,\label{si-eq: w-ver1}
\end{align}
where
\begin{align}
	w\left(\gamma^{-1}\Phi_{zy}^{2},\Phi_{zz},\beta,\delta\right)
	&:=\frac{4\beta^{2}\delta\left(\gamma^{-1}\Phi_{zy}^{2}\right)\left(\Phi_{zz}+1\right)^{2}}{\left[\left\{1+2\left(\Phi_{zz}+1\right)\beta\right\}\left(\gamma^{-1}\Phi_{zy}^{2}\right)+\left(\Phi_{zz}+1\right)^{2}\left(1+2\beta\right)\right]^{2}}\geq0. \label{si-eq: w-ver2}
\end{align}
Eq. (\ref{si-eq: w-ver1}) is calculated as follows: 
\begin{align}
	\Phi_{zz}=\frac{\sqrt{\left(w+1\right)^{2}\gamma^{-1}\Phi_{zy}^{2}+w^{2}}-1}{\left(w+1\right)}. 
\end{align}
From
\begin{align}
	\frac{\partial\Phi_{zz}}{\partial w}
	&=\frac{w+\sqrt{w^{2}+\left(w+1\right)^{2}\gamma^{-1}\Phi_{zy}^{2}}}{\left(w+1\right)^{2}\sqrt{w^{2}+\left(w+1\right)^{2}\gamma^{-1}\Phi_{zy}^{2}}}\geq0
\end{align}
and $w\geq0$, we obtain the following inequality: 
\begin{align}
	\underbrace{\sqrt{\gamma^{-1}\Phi_{zy}^{2}}-1}_{\partial J_{M}/\partial\Phi_{zz}=0}=\Phi_{zz}(w=0)\leq\Phi_{zz}(w)\leq\Phi_{zz}(w=\infty)=\underbrace{\sqrt{\gamma^{-1}\Phi_{zy}^{2}+1}}_{\partial J_{Q}/\partial\Phi_{zz}=0}.
\end{align}
This result shows that the values of $\Phi_{zz}$ satisfying $\partial J_{Q}/\partial \Phi_{zz}=0$ and $\partial J_{M}/\partial \Phi_{zz}=0$ are the upper and lower bounds, respectively, for the value of $\Phi_{zz}$ satisfying $\partial J/\partial \Phi_{zz}=0$. 

\subsection{Analysis of the Intersection between $\partial J/\partial \Phi_{zy}=0$ and $\partial J_{Q}/\partial \Phi_{zz}=0$}\label{si-subsec: Phizz-Q}
\begin{figure*}
	\includegraphics[width=85mm]{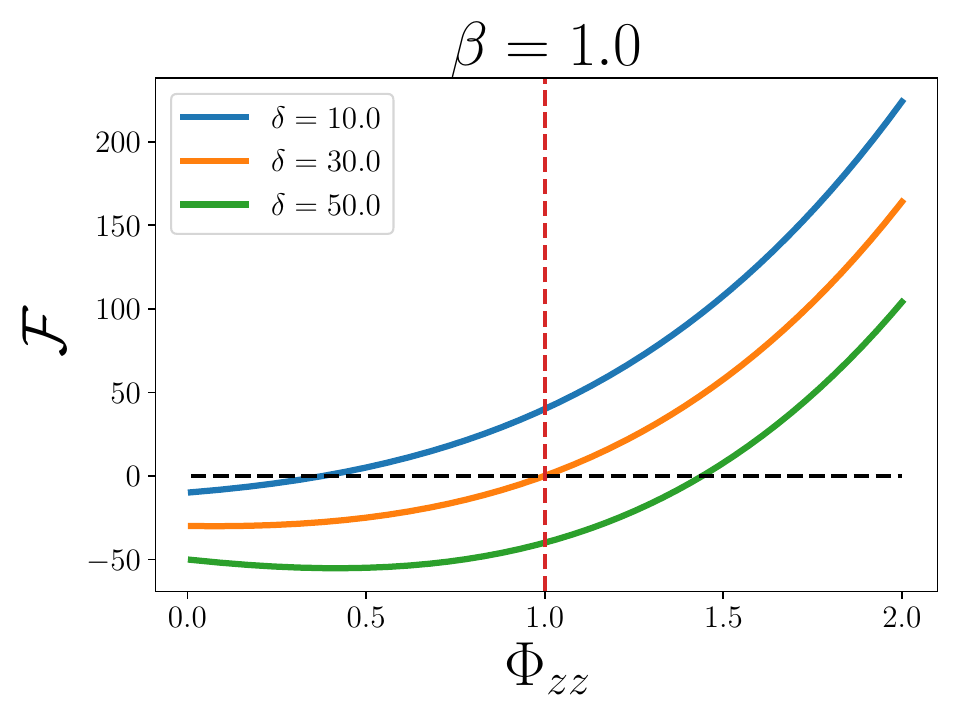}
	\caption{\label{fig-f-JQ}
	$\mcal{F}$ as a functions of $\Phi_{zz}$ [Eq. (\ref{si-eq: mcal-F})]. 
	Black and red dashed lines indicate $\mcal{F}=0$ and $\Phi_{zz}=1$, respectively. 
	}
\end{figure*}

In this subsection, we analyze the intersection between $\partial J/\partial \Phi_{zy}=0$ and $\partial J_{Q}/\partial \Phi_{zz}=0$. 
Since $\Phi_{zz}$ satisfying $\partial J_{Q}/\partial \Phi_{zz}=0$ is the upper bound for $\Phi_{zz}$ that satisfies $\partial J/\partial \Phi_{zz}=0$, the condition under which $\partial J_{Q}/\partial \Phi_{zz}=0$ intersects with $\partial J/\partial \Phi_{zy}=0$ gives a necessary condition where $\partial J/\partial \Phi_{zz}=0$ intersects with it. 

From $\partial J/\partial \Phi_{zy}=0$ [Eq. (\ref{si-eq: Phizy-star-ver2})] and $\partial J_{Q}/\partial \Phi_{zz}=0$ [Eq. (\ref{si-eq: pJpPhizz-ver2-Q})], we obtain the following equation: 
\begin{align}
	\left(\Phi_{zz}+1\right)\left(\Phi_{zz}-1\right)
	=\frac{(\Phi_{zz}+1)^{2}(1+2\beta)}{\left\{1+2(\Phi_{zz}+1)\beta\right\}}\left(\sqrt{\frac{4\Phi_{zz}\beta^{2}}{(\Phi_{zz}+1)\left\{1+2(\Phi_{zz}+1)\beta\right\}(1+2\beta)^{2}}\delta}-1\right). 
\end{align}
Rearranging this equation, we obtain the following equation: 
\begin{align}
	\mcal{F}(\Phi_{zz})=0, 
\end{align}
where
\begin{align}
	\mcal{F}(\Phi_{zz})
	&:=2\beta^{3}\Phi_{zz}^{4}+\beta^{2}\left(5+6\beta\right)\Phi_{zz}^{3}
	+2\beta\left(1+\beta\right)\left(2+3\beta\right)\Phi_{zz}^{2}+\left\{\left(1+\beta\right)^{2}\left(1+2\beta\right)-\beta^{2}\delta\right\}\Phi_{zz}-\beta^{2}\delta.\label{si-eq: mcal-F}
\end{align}
The derivatives of $\mcal{F}(\Phi_{zz})$ are given by
\begin{align}
	\frac{\partial \mcal{F}(\Phi_{zz})}{\partial \Phi_{zz}}
	&=8\beta^{3}\Phi_{zz}^{3}+3\beta^{2}\left(5+6\beta\right)\Phi_{zz}^{2}
	+4\beta\left(1+\beta\right)\left(2+3\beta\right)\Phi_{zz}+\left\{\left(1+\beta\right)^{2}\left(1+2\beta\right)-\beta^{2}\delta\right\},\\
	\frac{\partial^{2} \mcal{F}(\Phi_{zz})}{\partial \Phi_{zz}^{2}}
	&=24\beta^{3}\Phi_{zz}^{2}+6\beta^{2}\left(5+6\beta\right)\Phi_{zz}+4\beta\left(1+\beta\right)\left(2+3\beta\right),\\
	\frac{\partial^{3} \mcal{F}(\Phi_{zz})}{\partial \Phi_{zz}^{3}}
	&=48\beta^{3}\Phi_{zz}+6\beta^{2}\left(5+6\beta\right),\\
	\frac{\partial^{4} \mcal{F}(\Phi_{zz})}{\partial \Phi_{zz}^{4}}
	&=48\beta^{3}. 
\end{align}
$\partial^{2}\mcal{F}(\Phi_{zz})/\partial \Phi_{zz}^{2}>0$, $\partial^{3}\mcal{F}(\Phi_{zz})/\partial \Phi_{zz}^{3}>0$, and $\partial^{4}\mcal{F}(\Phi_{zz})/\partial \Phi_{zz}^{4}>0$ hold for $\Phi_{zz}\geq0$. 
Moreover, $\mcal{F}(0)=-\beta^{2}\delta<0$. 
These results indicate that $\mcal{F}(\Phi_{zz})=0$ always has a positive root [Fig. \ref{fig-f-JQ}]. 
Since $\Phi_{zy}^{2}=\gamma(\Phi_{zz}^{2}-1)$ [Eq. (\ref{si-eq: pJpPhizz-ver2-Q})], $\Phi_{zz}\geq1$ is required to ensure the nonnegativity of $\Phi_{zy}^{2}$. 
This condition is satisfied when $\mcal{F}(1)\leq0$, where
\begin{align}
	\mcal{F}(1)
	&=\left(1+4\beta\right)\left(1+2\beta\right)^{2}-2\beta^{2}\delta.
\end{align}
Whether $\mcal{F}(1)\leq0$ holds or not is determined by the following discriminant: 
\begin{align}
	\Theta_{Q}:=\frac{2\beta^{2}\delta}{\left(1+2\beta\right)^{2}\left(1+4\beta\right)}.\label{si-eq: Theta-Q-ver1}
\end{align}
When $\Theta_{Q}\geq1$, $\mcal{F}(1)\leq 0$ is satisfied, and $\partial J/\partial \Phi_{zy}=0$ and $\partial J_{Q}/\partial \Phi_{zz}=0$ intersect. 
Since $\Phi_{zz}$ satisfying $\partial J_{Q}/\partial \Phi_{zz}=0$ is the upper bound for $\Phi_{zz}$ that satisfies $\partial J/\partial \Phi_{zz}=0$, $\partial J_{Q}/\partial \Phi_{zz}=0$ always intersects with $\partial J/\partial \Phi_{zy}=0$ when $\partial J/\partial \Phi_{zz}=0$ intersects with it. 
Thus, $\Theta_{Q}\geq1$ gives a necessary condition where $\partial J/\partial \Phi_{zy}=0$ and $\partial J/\partial \Phi_{zz}=0$ intersect. 
When $\Theta_{Q}<1$, $\partial J/\partial \Phi_{zy}=0$ and $\partial J/\partial \Phi_{zz}=0$ do not intersect and nonzero memory control gains do not emerge. 

From $\beta:=E/D$ and $\delta:=QD/MF$, $\Theta_{Q}$ is rewritten as follows: 
\begin{align}
	\Theta_{Q}:=\frac{2D^{2}E^{2}}{\left(D+2E\right)^{2}\left(D+4E\right)}\frac{Q}{MF}.\label{si-eq: Theta-Q-ver2}
\end{align}
This expression explains the nonmonotonic phase transitions with respect to $D$ and $E$, and also confirms the scaling relation $Q/MF$. 

\subsection{Analysis of the Intersection between $\partial J/\partial \Phi_{zy}=0$ and $\partial J_{M}/\partial \Phi_{zz}=0$}\label{si-subsec: Phizz-M}
\begin{figure*}
	\includegraphics[width=170mm]{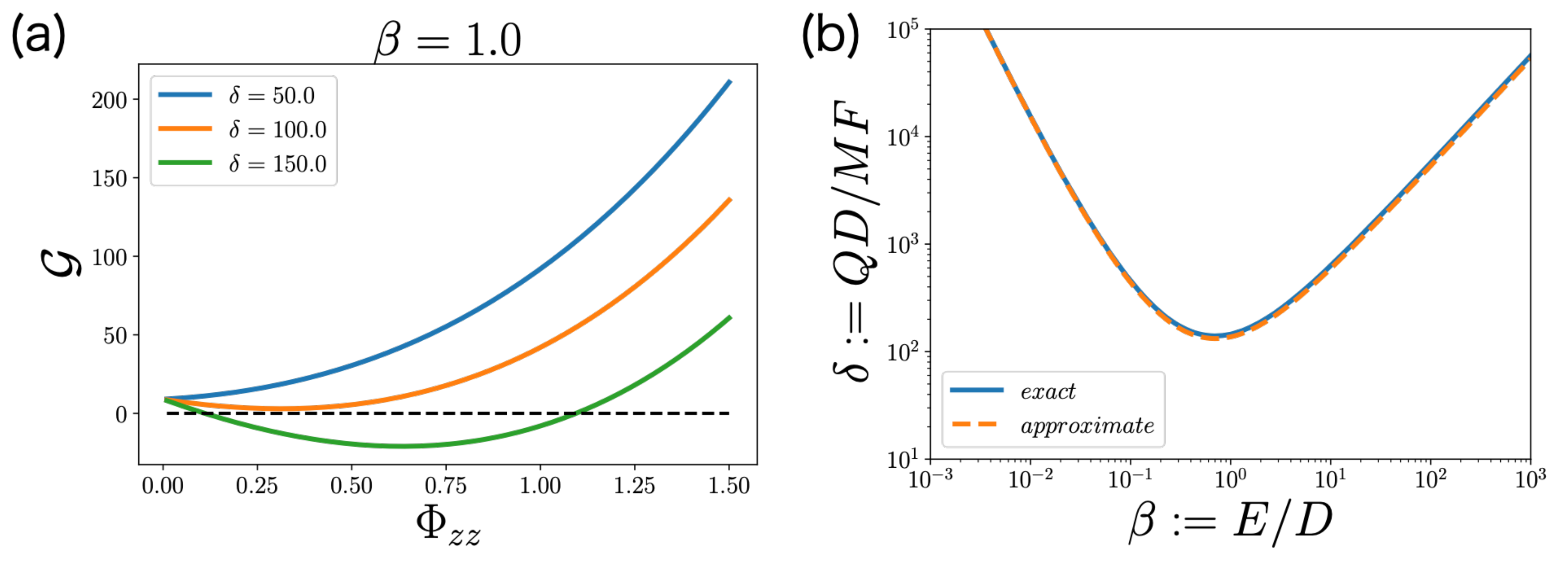}
	\caption{\label{fig-f-JM}
	(a) $\mcal{G}$ as a functions of $\Phi_{zz}$ [Eq. (\ref{si-eq: mcal-G})]. 
	Black dashed line indicates $\mcal{G}=0$. 
	(b) Blue and orange curves are the parameters where $\mcal{G}(\Phi_{zz})=0$ [Eq. (\ref{si-eq: mcal-G})] and $\tilde{\mcal{G}}(\Phi_{zz})=0$ [Eq. (\ref{si-eq: G-approx})] have a positive root, respectively. 
	While the blue curve is obtained numerically, the orange curve is given by $\Theta_{M}=1$ [Eq. (\ref{si-eq: Theta-M})]. 
	}
\end{figure*}

In this subsection, we analyze the intersection between $\partial J/\partial \Phi_{zy}=0$ and $\partial J_{M}/\partial \Phi_{zz}=0$. 
Since $\Phi_{zz}$ satisfying $\partial J_{M}/\partial \Phi_{zz}=0$ is the lower bound for $\Phi_{zz}$ that satisfies $\partial J/\partial \Phi_{zz}=0$, the condition under which $\partial J_{M}/\partial \Phi_{zz}=0$ intersects with $\partial J/\partial \Phi_{zy}=0$ gives a sufficient condition where $\partial J/\partial \Phi_{zz}=0$ intersects with it. 

From $\partial J/\partial \Phi_{zy}=0$ [Eq. (\ref{si-eq: Phizy-star-ver2})] and $\partial J_{M}/\partial \Phi_{zz}=0$ [Eq. (\ref{si-eq: pJpPhizz-ver2-M})], we obtain the following equation: 
\begin{align}
	\Phi_{zy}^{2}=\gamma\frac{(\Phi_{zz}+1)^{2}(1+2\beta)}{\left\{1+2(\Phi_{zz}+1)\beta\right\}}\left(\sqrt{\frac{4\Phi_{zz}\beta^{2}}{(\Phi_{zz}+1)\left\{1+2(\Phi_{zz}+1)\beta\right\}(1+2\beta)^{2}}\delta}-1\right).
\end{align}
Rearranging this equation, we obtain the following equation: 
\begin{align}
	\mcal{G}(\Phi_{zz})=0, 
\end{align}
where
\begin{align}
	\mcal{G}(\Phi_{zz})
	&:=2\beta^{3}\Phi_{zz}^{4}+\beta^{2}\left(5+12\beta\right)\Phi_{zz}^{3}\nonumber\\
	&\ \ \ +\beta\left(1+2\beta\right)\left(4+13\beta\right)\Phi_{zz}^{2}+\left\{\left(1+2\beta\right)^{2}\left(1+6\beta\right)-\beta^{2}\delta\right\}\Phi_{zz}+\left(1+2\beta\right)^{3}.\label{si-eq: mcal-G}
\end{align}
The derivatives of $\mcal{G}(\Phi_{zz})$ are given by 
\begin{align}
	\frac{\partial \mcal{G}(\Phi_{zz})}{\partial \Phi_{zz}}
	&=8\beta^{3}\Phi_{zz}^{3}+3\beta^{2}\left(5+12\beta\right)\Phi_{zz}^{2}
	+2\beta\left(1+2\beta\right)\left(4+13\beta\right)\Phi_{zz}+\left\{\left(1+2\beta\right)^{2}\left(1+6\beta\right)-\beta^{2}\delta\right\},\\
	\frac{\partial^{2} \mcal{G}(\Phi_{zz})}{\partial \Phi_{zz}^{2}}
	&=24\beta^{3}\Phi_{zz}^{2}+6\beta^{2}\left(5+12\beta\right)\Phi_{zz}+2\beta\left(1+2\beta\right)\left(4+13\beta\right),\\
	\frac{\partial^{3} \mcal{G}(\Phi_{zz})}{\partial \Phi_{zz}^{3}}
	&=48\beta^{3}\Phi_{zz}+6\beta^{2}\left(5+12\beta\right),\\
	\frac{\partial^{4} \mcal{G}(\Phi_{zz})}{\partial \Phi_{zz}^{4}}
	&=48\beta^{3}.
\end{align}
$\partial^{2}\mcal{G}(\Phi_{zz})/\partial \Phi_{zz}^{2}>0$, $\partial^{3}\mcal{G}(\Phi_{zz})/\partial \Phi_{zz}^{3}>0$, and $\partial^{4}\mcal{G}(\Phi_{zz})/\partial \Phi_{zz}^{4}>0$ hold for $\Phi_{zz}\geq0$. 
Moreover, $\mcal{G}(0)=\left(1+2\beta\right)^{3}>0$. 
These results indicate that whether $\mcal{G}(\Phi_{zz})=0$ has positive roots or not depends on $\partial\mcal{G}(\Phi_{zz})/\partial \Phi_{zz}$:  At least, when $\partial\mcal{G}(\Phi_{zz})/\partial \Phi_{zz}\geq0$, $\mcal{G}(\Phi_{zz})=0$ does not have positive roots [Fig. \ref{fig-f-JM}(a)]. 
However, a direct analysis of $\mcal{G}(\Phi_{zz})=0$ is challenging because $\mcal{G}(\Phi_{zz})$ is a fourth-order function. 
To address this issue, we approximate $\mcal{G}(\Phi_{zz})$ by a second-order expansion around the origin as follows: 
\begin{align}
	\mcal{G}(\Phi_{zz})
	\approx
	\tilde{\mcal{G}}(\Phi_{zz})
	&:=\mcal{G}(0)+\frac{\partial \mcal{G}(0)}{\partial \Phi_{zz}}\Phi_{zz}+\frac{1}{2}\frac{\partial^{2} \mcal{G}(0)}{\partial \Phi_{zz}^{2}}\Phi_{zz}^{2}\nonumber\\
	&=\left(1+2\beta\right)^{3}+\left\{\left(1+2\beta\right)^{2}\left(1+6\beta\right)-\beta^{2}\delta\right\}\Phi_{zz}+\beta\left(1+2\beta\right)\left(4+13\beta\right)\Phi_{zz}^{2}.
	\label{si-eq: G-approx}
\end{align}
The solutions of $\tilde{\mcal{G}}(\Phi_{zz})=0$ are given by 
\begin{align}
	\Phi_{zz}
	&=\frac{\left\{\beta^{2}\delta-\left(1+2\beta\right)^{2}\left(1+6\beta\right)\right\}\pm\sqrt{\left\{\beta^{2}\delta-\left(1+2\beta\right)^{2}\left(1+6\beta\right)\right\}^{2}-4\beta\left(1+2\beta\right)^{4}\left(4+13\beta\right)}}{2\beta\left(1+2\beta\right)\left(4+13\beta\right)}.\label{si-eq: JMsec-Phizz-ver1}
\end{align}
To ensure that $\Phi_{zz}$ is real, the following condition must hold: 
\begin{align}
	&\left\{\beta^{2}\delta-\left(1+2\beta\right)^{2}\left(1+6\beta\right)\right\}^{2}-4\beta\left(1+2\beta\right)^{4}\left(4+13\beta\right)\geq0\nonumber\\
	\Rightarrow\ 
	&\beta^{4}\delta^{2}-2\beta^{2}\left(1+2\beta\right)^{2}\left(1+6\beta\right)\delta+\left(1+2\beta\right)^{4}\left(1-4\beta-16\beta^{2}\right)\geq0.\label{si-eq: JMsec-condition-ver1}
\end{align}
This equality is satisfied when 
\begin{align}
	\delta
	&=\frac{\beta^{2}\left(1+2\beta\right)^{2}\left(1+6\beta\right)\pm\sqrt{\beta^{4}\left(1+2\beta\right)^{4}\left(1+6\beta\right)^{2}-\beta^{4}\left(1+2\beta\right)^{4}\left(1-4\beta-16\beta^{2}\right)}}{\beta^{4}}\nonumber\\
	&=\frac{\left(1+2\beta\right)^{2}\left\{\left(1+6\beta\right)\pm\sqrt{4\beta\left(4+13\beta\right)}\right\}}{\beta^{2}}.\label{si-eq: JMsec-condition-ver2}
\end{align}
Thus, Eq. (\ref{si-eq: JMsec-condition-ver1}) becomes
\begin{align}
	\delta\leq\frac{\left(1+2\beta\right)^{2}\left\{\left(1+6\beta\right)-\sqrt{4\beta\left(4+13\beta\right)}\right\}}{\beta^{2}},\quad
	\frac{\left(1+2\beta\right)^{2}\left\{\left(1+6\beta\right)+\sqrt{4\beta\left(4+13\beta\right)}\right\}}{\beta^{2}}\leq\delta. \label{si-eq: JMsec-condition-ver3}
\end{align}
Furthermore, to ensure that $\Phi_{zz}$ is nonnegative, the following condition must hold: 
\begin{align}
	\beta^{2}\delta-\left(1+2\beta\right)^{2}\left(1+6\beta\right)\geq0
	\ \Rightarrow\ 
	\delta\geq\frac{\left(1+2\beta\right)^{2}\left(1+6\beta\right)}{\beta^{2}}.\label{si-eq: JMsec-condition-ver4}
\end{align}
From Eqs. (\ref{si-eq: JMsec-condition-ver3}) and (\ref{si-eq: JMsec-condition-ver4}), the condition under which $\tilde{\mcal{G}}(\Phi_{zz})=0$ has positive solutions is given by
\begin{align}
	\delta\geq\frac{\left(1+2\beta\right)^{2}\left\{\left(1+6\beta\right)+\sqrt{4\beta\left(4+13\beta\right)}\right\}}{\beta^{2}}. 
\end{align}
As a result, the condition under which $\partial J/\partial \Phi_{zy}=0$ and $\partial J_{M}/\partial \Phi_{zz}=0$ intersect is approximately given by the following discriminant: 
\begin{align}
	\Theta_{M}:=\frac{\beta^{2}\delta}{\left(1+2\beta\right)^{2}\left\{\left(1+6\beta\right)+\sqrt{4\beta\left(4+13\beta\right)}\right\}}.\label{si-eq: Theta-M}
\end{align}
When $\Theta_{M}\geq1$, $\mcal{G}(\Phi_{zz})\approx\tilde{\mcal{G}}(\Phi_{zz})=0$ has positive roots, and $\partial J_{M}/\partial \Phi_{zz}=0$ intersects with $\partial J/\partial \Phi_{zy}=0$. 
Since $\Phi_{zz}$ satisfying $\partial J_{M}/\partial \Phi_{zz}=0$ is the lower bound for $\Phi_{zz}$ that satisfies $\partial J/\partial \Phi_{zz}=0$, $\partial J/\partial \Phi_{zz}=0$ always intersects with $\partial J/\partial \Phi_{zy}=0$ when $\partial J_{M}/\partial \Phi_{zz}=0$ intersects with it. 
Thus, $\Theta_{M}\geq1$ gives a sufficient condition where $\partial J/\partial \Phi_{zy}=0$ and $\partial J/\partial \Phi_{zz}=0$ intersect. 
When $\Theta_{M}\geq1$, $\partial J/\partial \Phi_{zy}=0$ and $\partial J/\partial \Phi_{zz}=0$ intersect and nonzero memory control gains emerge. 

We note that $\Theta_{M}$ is an approximate discriminant because the forth-order equation $\mcal{G}(\Phi_{zz})=0$ is approximated with the second-order equation $\tilde{\mcal{G}}(\Phi_{zz})=0$. 
Nevertheless, the approximate discriminant $\Theta_{M}$ is highly consistent with the exact one [Fig. \ref{fig-f-JM}(b)]. 
Therefore, the approximate equation $\tilde{\mcal{G}}(\Phi_{zz})=0$ captures the behavior of the exact equation $\mcal{G}(\Phi_{zz})=0$. 

Moreover, $\Theta_{Q}>\Theta_{M}$ always holds. 
It is consistent with the fact that the values of $\Phi_{zz}$ satisfying $\partial J_{Q}/\partial \Phi_{zz}=0$ and $\partial J_{M}/\partial \Phi_{zz}=0$ serve as  the upper and lower bounds, respectively, for the value of $\Phi_{zz}$ that satisfies $\partial J/\partial \Phi_{zz}=0$. 

From $\beta:=E/D$ and $\delta:=QD/MF$, $\Theta_{M}$ is rewritten as follows: 
\begin{align}
	\Theta_{M}:=\frac{D^{2}E^{2}}{\left(D+2E\right)^{2}\left\{\left(D+6E\right)+\sqrt{4E\left(4D+13E\right)}\right\}}\frac{Q}{MF}.
\end{align}
This expression explains the nonmonotonic phase transitions with respect to $D$ and $E$, and also confirms the scaling relation $Q/MF$. 

\section{Phase Transition with respect to Environmental Volatility}
Recent human experiments have reported that while increasing sensory uncertainty induces a nonmonotonic phase transition between memoryless and memory-based estimation strategies, increasing environmental volatility induces a monotonic transition from memory-based to memoryless estimation strategies \cite{tavoni_human_2022}. 
In the main text, we analytically demonstrated the nonmonotonic phase transition with respect to sensory uncertainty, which has also been reported numerically in our previous work \cite{tottori_resource_2025,tottori_theory_2025}. 
In contrast, it remains unclear whether our framework can capture the monotonic transition associated with environmental volatility. 
In this section, we analytically demonstrate this transition and confirm it through numerical experiments. 

To incorporate environmental volatility, we reformulate the environmental state dynamics as follows: 
\begin{align}
	dx_{t}=-\ve x_{t}dt+\sqrt{\ve D}d\omega_{t},
	\label{eq: state-mod}
\end{align}
where $\ve>0$. 
In the steady state, the autocorrelation function is given by 
\begin{align}
	\mb{E}\left[x_{t}x_{t+\tau}\right]=\frac{D}{2}\exp(-\ve\tau).
\end{align}
Since larger $\ve$ accelerates the decay of the past environmental information, $\ve$ characterizes the environmental volatility. 
The rest of the formulation is the same as in the main text. 

In this problem setting, the steady-state equation of the covariance matrix is given by 
\begin{align}
	\left(\begin{array}{cc}
		0&0\\
		0&0\\
	\end{array}\right)
	&=\left(\begin{array}{cc}
		\ve D&0\\
		0&F\\
	\end{array}\right)
	+\left(\begin{array}{cc}
		-\ve&0\\
		-\Phi_{zy}&-\Phi_{zz}\\
	\end{array}\right)\left(\begin{array}{cc}
		\Sigma_{xx}&\Sigma_{xz}\\
		\Sigma_{zx}&\Sigma_{zz}\\
	\end{array}\right)+\left(\begin{array}{cc}
		\Sigma_{xx}&\Sigma_{xz}\\
		\Sigma_{zx}&\Sigma_{zz}\\
	\end{array}\right)\left(\begin{array}{cc}
		-\ve&-\Phi_{zy}\\
		0&-\Phi_{zz}\\
	\end{array}\right).
\end{align}
By multiplying $\ve^{-1}$, we obtain 
\begin{align}
	\left(\begin{array}{cc}
		0&0\\
		0&0\\
	\end{array}\right)
	&=\left(\begin{array}{cc}
		D&0\\
		0&\ve^{-1}F\\
	\end{array}\right)
	+\left(\begin{array}{cc}
		-1&0\\
		-\ve^{-1}\Phi_{zy}&-\ve^{-1}\Phi_{zz}\\
	\end{array}\right)\left(\begin{array}{cc}
		\Sigma_{xx}&\Sigma_{xz}\\
		\Sigma_{zx}&\Sigma_{zz}\\
	\end{array}\right)+\left(\begin{array}{cc}
		\Sigma_{xx}&\Sigma_{xz}\\
		\Sigma_{zx}&\Sigma_{zz}\\
	\end{array}\right)\left(\begin{array}{cc}
		-1&-\ve^{-1}\Phi_{zy}\\
		0&-\ve^{-1}\Phi_{zz}\\
	\end{array}\right).
\end{align}
In addition, the objective function can be calculated as follows: 
\begin{align}
	J[\hat{x},v]
	&=\lim_{T \to \infty} \frac{1}{T} \mathbb{E} \left[ \int_0^T \left( Q(x_{t} - \hat{x}_{t})^2 + M v_{t}^{2} \right) dt \right]\nonumber\\
	&=\lim_{T \to \infty} \frac{1}{T} \mathbb{E} \left[ \int_0^T \left( Q(x_{t} - \hat{x}_{t})^2 + M \left(-\Phi_{zy}y-\Phi_{zz}z\right)^{2} \right) dt \right]\nonumber\\
	&=\lim_{T \to \infty} \frac{1}{T} \mathbb{E} \left[ \int_0^T \left( Q(x_{t} - \hat{x}_{t})^2 + \ve^{2}M \left(-\ve^{-1}\Phi_{zy}y-\ve^{-1}\Phi_{zz}z\right)^{2} \right) dt \right].
\end{align}
Therefore, by replacing $F$, $M$, $\Phi_{zy}$, and $\Phi_{zz}$ with $F':=\ve^{-1}F$, $M':=\ve^{2}M$, $\Phi_{zy}':=\ve^{-1}\Phi_{zy}$, and $\Phi_{zz}':=\ve^{-1}\Phi_{zz}$, respectively, the same discussion is possible. 
Under this replacement, $\beta:=E/D$ remains the same, whereas $\delta:=QD/MF$ changes as
\begin{align}
	\delta'=\frac{Q'}{M'}\frac{D'}{F'}
	=\frac{1}{\ve}\frac{Q}{M}\frac{D}{F}
	=\frac{\delta}{\ve}. 
\end{align}
Therefore, increasing environmental volatility $\ve$ is the same as decreasing energy availability or memory accuracy $\delta$, which explains the monotonic phase transition from memory-based to memoryless estimation strategies  \cite{tavoni_human_2022}. 

\begin{figure}
	\includegraphics[width=85mm]{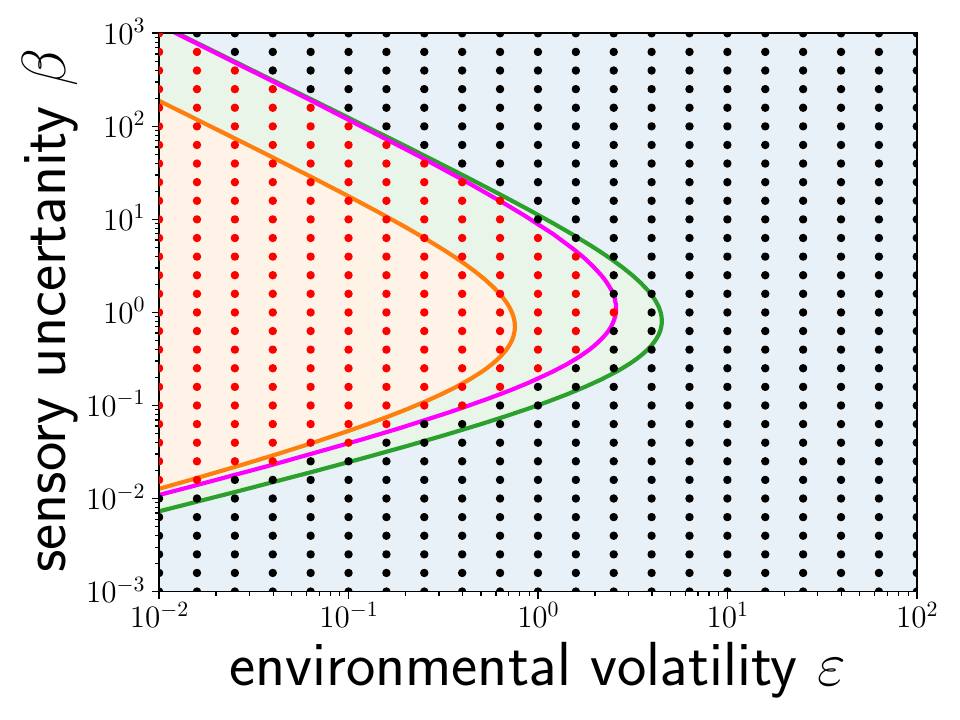}
	\caption{\label{fig-volatility}
	Phase transitions with respect to environmental volatility $\ve$ and sensory uncertainty $\beta$. 
	Black and red dots indicate the memoryless and memory-based estimation strategies, respectively. 
	These dots are obtained by solving the observation-based Riccati equation numerically. 
	Magenta, green, and orange curves represent the parameter values at which $\partial J/\partial \Phi_{zz}=0$, $\partial J_{Q}/\partial \Phi_{zz}=0$, and $\partial J_{M}/\partial \Phi_{zz}=0$ intersect tangentially with $\partial J/\partial \Phi_{zy}=0$, respectively. 
	The magenta curve ($\Theta_{T}=1$) is obtained numerically, whereas the green ($\Theta_{Q}=1$) and orange ($\Theta_{M}=1$) curves are obtained analytically from Eqs. (\ref{si-eq: Theta-Q-ve}) and (\ref{si-eq: Theta-M-ve}), respectively. 
	Blue, green, and orange regions correspond to $\Theta_{Q}<1$, $1\leq\Theta_{Q}$ and $\Theta_{M}<1$, and $1\leq\Theta_{M}$, respectively. 
	In the numerical calculations, $\ve$ and $E$ vary, while the other parameters are fixed at $D=1.0$, $F=0.01$, $Q=1.0$, and $M=1.0$. 
	}
\end{figure}

We verify these analytical insights by numerically solving the observation-based Riccati equation [Eqs. (\ref{si-eq: ODE of Si}) and (\ref{si-eq: ODE of Pi})]. 
The numerical procedure to solve the observation-based Riccati equation follows the same method as in our previous work \cite{tottori_resource_2025,tottori_theory_2025}. 
The result is shown in Fig. \ref{fig-volatility}. 
Figure \ref{fig-volatility} shows that while sensory uncertainty $\beta$ induces a nonmontonic phase transition between memoryless and memory-based estimation strategies, environmental volatility $\ve$ induces a monotonic transition from memory-based to memoryless estimation strategies. 
Furthermore, in this problem setting, the analytical discriminants $\Theta_{Q}$ and $\Theta_{M}$ are modified as follows: 
\begin{align}
	\Theta_{Q}&=\frac{2\beta^{2}}{\left(1+2\beta\right)^{2}\left(1+4\beta\right)}\frac{\delta}{\ve},\label{si-eq: Theta-Q-ve}\\
	\Theta_{M}&=\frac{\beta^{2}}{\left(1+2\beta\right)^{2}\left\{\left(1+6\beta\right)+\sqrt{4\beta\left(4+13\beta\right)}\right\}}\frac{\delta}{\ve}.\label{si-eq: Theta-M-ve}
\end{align}
When $\Theta_{Q}<1$ [Fig. \ref{fig-volatility}, blue region], only memoryless estimation strategies appear [Fig. \ref{fig-volatility}, black dots], whereas when $\Theta_{M}\geq1$ [Fig. \ref{fig-volatility}, orange region], only memory-based estimation strategies appear [Fig. \ref{fig-volatility}, red dots]. 
Therefore, our analytical discriminants remain valid even when environmental volatility $\ve$ varies.

\bibliographystyle{apsrev4-2}
\nocite{apsrev42Control}
\bibliography{251015_OMSC_ref_cleaned}